\documentclass{nature}
\usepackage{graphicx}
\graphicspath{{figures/}{./}}
\makeatletter
\let\saved@includegraphics\includegraphics
\AtBeginDocument{\let\includegraphics\saved@includegraphics}
\renewenvironment*{figure}{\@float{figure}}{\end@float}
\makeatother

\usepackage{multirow}
\usepackage{subfigure}
\usepackage{mathtools}
\usepackage[colorlinks=true,linkcolor=blue,citecolor=black,urlcolor=blue]{hyperref}
\usepackage{authblk}
\usepackage[labelfont=bf]{caption}
\usepackage[figurename=Figure]{caption}
\usepackage{siunitx}
\usepackage[dvipsnames]{xcolor}
\usepackage{soul}
\usepackage{xfrac}
\usepackage{amsmath}
\usepackage{amssymb}
\usepackage{soul,xcolor}
\usepackage{lineno}
\usepackage{amsfonts}
\usepackage{booktabs}%for table double line
\usepackage{tabularx}%for table size
\usepackage{adjustbox}
\usepackage[utf8]{inputenc}
\usepackage{graphicx}
\DeclareGraphicsExtensions{.eps}
\usepackage{lipsum}
\usepackage{bm}
\usepackage{siunitx}
\usepackage{xparse}
\usepackage{textcomp} % This package is just to give the text quote '
\usepackage{enumitem}
\usepackage{amsmath,amssymb}
	\usepackage{color}
	\usepackage{xcolor}
	\usepackage{longtable}
	\usepackage{footnote}
	\usepackage{physics}
	\usepackage{soul}
	\usepackage{dcolumn}
	\usepackage{array}
	\usepackage{blindtext, rotating}
	\usepackage{braket}
	\usepackage{multirow}
	\setlist{noitemsep,leftmargin=*}
	\usepackage{gensymb}

	\usepackage{hyperref}
	\hypersetup{
		colorlinks=true,
		urlcolor= blue,
		citecolor=blue,
		linkcolor= blue,
		bookmarksopen=false,
	}

\usepackage{booktabs}\usepackage{placeins}
\newcommand{\nn}{\hat{\mathbf n}}

\newcommand{\zz}{\hat{\mathbf z}}
\newcommand{\R}{\mathbf R}
\newcommand{\rr}{\mathbf r}
\begin{document}
\setstcolor{red}
\title{Exciton design via topological spin textures}
\author[1,2$\dag$]{Karim Rezouali}
\author[3,1,4$\dag\dag$]{Selcuk S\"{o}zeri}
\author[3,*]{Samir Lounis}
\affil[1]{Faculty of Physics and CENIDE, University of Duisburg-Essen, D-47053 Duisburg, Germany}
\affil[2]{Laboratoire de Physique Th\'eorique, Facult\'e des Sciences Exactes, Universit\'e de Bejaia, 06000 Bejaia, Alg\'erie}
\affil[3]{Institute of Physics and Halle-Berlin-Regensburg Cluster of Excellence CCE, Martin-Luther-University Halle-Wittenberg, 06099 Halle (Saale), Germany}
\affil[4]{Peter Grünberg Institut, Forschungszentrum Jülich and JARA, D-52425 Jülich, Germany}
\affil[$\dag$]{karim.rezouali@univ-bejaia.dz}
\affil[$\dag\dag$]{selcuk.soezeri@physik.uni-halle.de}
\affil[*]{samir.lounis@physik.uni-halle.de}

\maketitle
\begin{abstract}
Excitons carry energy and information through a semiconductor without carrying charge. This makes them attractive for optoelectronics, but also hard to guide. Magnetic skyrmions, on the other hand, are moved by small electrical currents. Here we develop a theory of excitons interacting with a skyrmion in which the electron and the hole are treated as distinct constituents, each with its own spin--orbit coupling. We find that the topology of the skyrmion is transferred to the exciton only when the exchange coupling exceeds the relevant spin--orbit splitting. The exciton then experiences a full flux quantum, with opposite signs in the two valleys. The quantum metric of the spin texture creates a ring-shaped barrier that traps excitons, whereas the topological charge fixes the angular momentum of the trapped states and hence their optical selection rules. Because the trap moves with the skyrmion, excitons can be transported by current pulses, and small skyrmions accommodate a single exciton, which opens the door for potential hybrid applications in future information technology intertwining topological magnetic bits with optical excitations.
\end{abstract}

\section*{Introduction}
Magnetic skyrmions are localized entities consisting of  topologically non-trivial spin swirling textures in which the magnetization whirls around a reversed core~\cite{bogdanov1989,bogdanov1994,bogdanov1994-2,rossler2011}. They are stabilized by the Dzyaloshinskii--Moriya interaction~\cite{dzyaloshinsky1958,moriya1960} and have been observed on length scales from micrometres down to a few nanometres~\cite{muhlbauer2009,heinze2011,romming2015, wiesendanger2016,fert2017}.
%\samircom{Add a review, maybe Fert, Wiesendanger?}. 
Individual skyrmions can be written and deleted with a spin-polarized tip~\cite{romming2013}, and they are set in motion by electrical currents of modest density~\cite{nagaosa2013,fert2013}. This mobility is at the origin of the racetrack concept~\cite{parkin2008}, in which information is encoded in magnetic objects that are shifted along a track by current pulses.

Skyrmions do not only carry information themselves; they also act on particles passing through them. An electron whose spin follows the texture adiabatically acquires a Berry phase, which is equivalent to an emergent magnetic field and gives rise to the topological Hall effect~\cite{schulz2012,bruno2004,everschor-Sitte2014,verma2022}. Further magneto-transport effects, associated to chirality, non-collinearity and spin-mixing can emerge such as the  chiral and non-collinear Hall effects~\cite{lux2020,bouaziz2021}
%\samircom{cite Fabian Lux paper, and paper of Juba}
as well as a whole family of spin-mixing magnetoresistances \cite{crum2015,hanneken2015,fernandes2022}.
%\samircom{cite paper of Dax Crum, and Imara Lima Fernandes, and the nature nanotechnology of Wiesendanger (non-collinear magnetoresistance)}.
Even when the electron spin cannot follow the texture, the scattering remains asymmetric~\cite{denisov2016}. Similar effects have been predicted for magnons~\cite{schutte2014,zhang2017,pan2024,seungho2023}. In all these cases the particle is elementary, and its coupling to the texture is characterized by a single exchange constant.

Excitons, the bound electron--hole pairs that dominate the optical response of two-dimensional semiconductors~\cite{cardona2010,combescot2016,knox1983}, are a natural next candidate, but they raise questions that do not arise for electrons or magnons. An exciton transports energy without net charge~\cite{mueller2018}, which is precisely why it is so difficult to control: neutral excitons are insensitive to electric fields, and guiding them has so far required spatially indirect excitons in electrostatic landscapes~\cite{high2008,winbow2011}, thermal gradients~\cite{cadiz2018} or surface acoustic waves~\cite{peng2022}. A skyrmion that could hold an exciton and carry it along a racetrack would provide a very different handle. Whether this is possible depends, however, on how an exciton perceives a spin texture in the first place, and here the composite nature of the exciton becomes essential. In transition-metal dichalcogenides, for instance, the spin of the hole is locked to the valley by an Ising spin--orbit splitting of a few hundred meV, while the spin of the electron is split by only a few meV. The two constituents of one and the same exciton therefore respond very differently to the texture, and it is not obvious what the exciton as a whole experiences.

\begin{figure}[!tbp]
\centering\includegraphics[width=\textwidth]{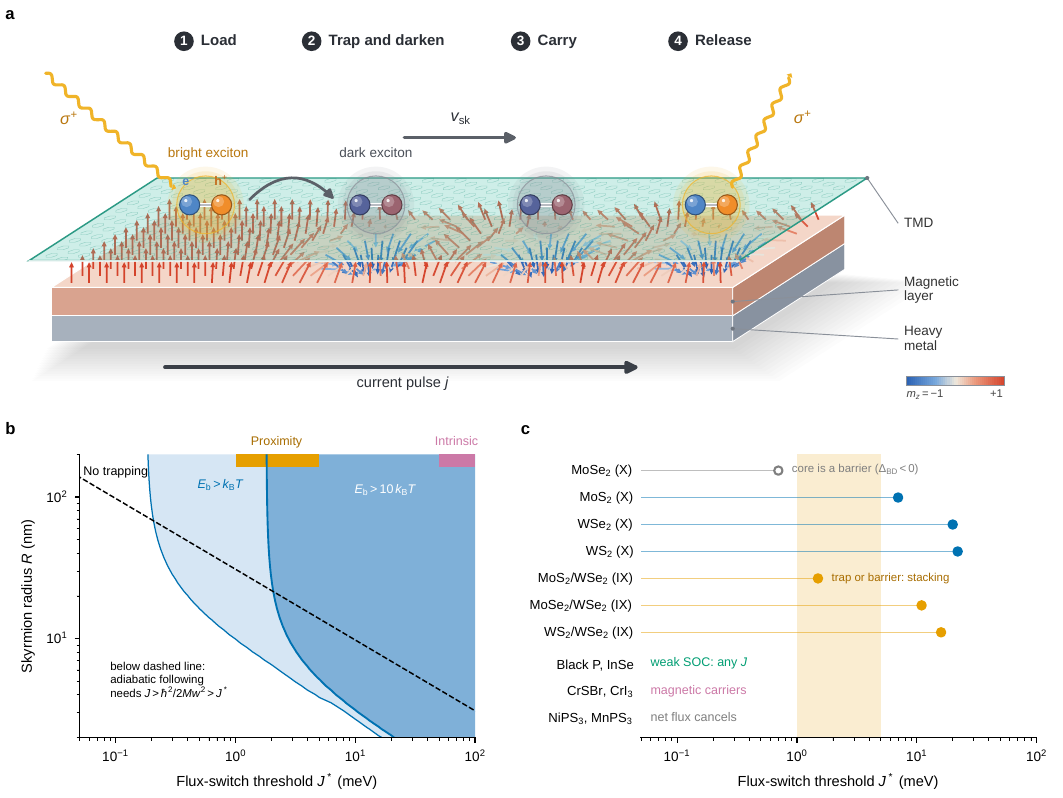}
\caption{\textbf{Excitons in a skyrmion: a map.} \textbf{a}, Illustration of the proposed scheme in a heterostructure of a two-dimensional semiconductor (TMD), an insulating skyrmion host and a heavy-metal layer that drives the skyrmions. (1) Circularly polarized light creates a bright exciton (electron e$^-$ and hole h$^+$). (2) In the skyrmion, bright and dark excitons are mixed and the exciton is held in the core. (3) A current pulse $j$ moves the skyrmion together with the exciton. (4) The exciton is released and recombines radiatively. The rendering is schematic and not to scale. \textbf{b}, Regimes as a function of the exchange coupling $J$ and the skyrmion radius $R$. Shaded areas: exciton trapping with binding energy above $k_BT$ and above $10\,k_BT$ (transport window) at 4.2\,K, for an exchange well of depth $2J$ and wall width $R/5$. Vertical lines: onset of adiabatic following for wall widths of 1.5 and 5\,nm. Dash-dotted lines: radii beyond which the trapped resonance lives longer than 10 and 100\,ps (walls of 1.5\,nm; profile dependent).
Bars on top: indicative exchange scales for proximity coupling (orange, 1--5\,meV)
and for magnetic semiconductors with magnetic carriers (purple, above 50\,meV).
\textbf{c}, Flux-switch threshold $J^*$ of two-dimensional excitonic materials: $|\Delta_{BD}|/2$ from measured bright--dark splittings for intralayer excitons (X)~\cite{robert2020}, and the conduction-band splitting $\Delta_c$ of the electron layer for interlayer excitons (IX)~\cite{kormanyos2015}; weak spin--orbit semiconductors, magnetic semiconductors and antiferromagnets are indicated qualitatively. See Supplementary Tables~\ref{tab:S1} and \ref{tab:S2}.}
\label{fig:map}
\end{figure}

In this work, we address this question by building the theory of an exciton in a skyrmion from its constituents up. We find that the emergent flux felt by the exciton is not guaranteed by topology alone. It appears only when the exchange coupling exceeds the relevant spin--orbit splitting, and it then takes opposite signs in the two valleys. Once the flux is present, the skyrmion acts as a trap for excitons. The two geometric ingredients of the problem play distinct roles in this trap: the quantum metric of the texture produces the confining barrier, while the Berry curvature, that is the topology, determines the angular momentum of the trapped states and therefore which light can address them. Since the trap is the skyrmion itself, it can be moved by a current, and we show that the exciton follows with negligible loss. The theory depends on only a few parameters, namely the exchange coupling, the spin--orbit splitting, the size of the skyrmion and the temperature. This allows us to summarize the different regimes in a single map (Fig.~\ref{fig:map}) and to discuss which of the known two-dimensional excitonic materials are suited to realize them.

\section*{Results}
\subsection*{Emergent flux of a composite exciton}
\begin{figure}[!tbp]
\centering\includegraphics[width=\textwidth]{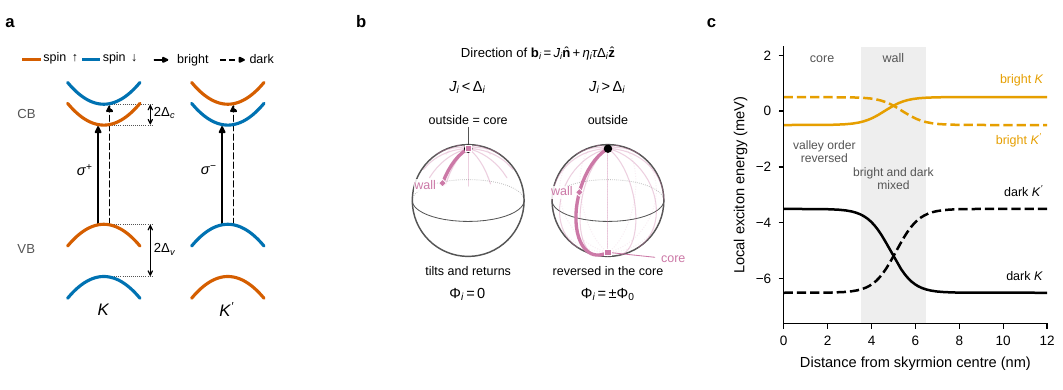}
\caption{\textbf{Bands, excitons and the flux switch.} \textbf{a}, Band edges of a transition-metal dichalcogenide at the $K$ and $K'$ valleys. The valence band has a large Ising splitting $2\Delta_v$ and the conduction band a small one, $2\Delta_c$, with the spin ordering reversed between the valleys. Solid arrows: bright, spin-conserving transitions addressed by $\sigma^\pm$ light; dashed arrows: spin-forbidden dark transitions. The band ordering shown is that of the Mo-based compounds; the bright--dark splitting of the excitons, $\Delta_\mathrm{BD}$, additionally contains the electron--hole exchange, which in MoS$_2$ places the dark exciton below the bright one~\cite{robert2020}. \textbf{b}, Path of the direction of the effective field $\mathbf b_i$ from outside the skyrmion to its core. For $J_i<\Delta_i$ the field only tilts in the wall and returns, and the emergent flux vanishes; for $J_i>\Delta_i$ it is reversed in the core, sweeps the full sphere and carries one flux quantum. \textbf{c}, Local energies of bright and dark excitons in the two valleys across a skyrmion (radius 5\,nm, wall 1\,nm), obtained from Eq.~\eqref{eq:Htau} (Methods) with $J_c=1$\,meV, $J_v=0.5$\,meV and $\Delta_\mathrm{BD}=5$\,meV. The valley ordering is reversed in the core, and bright and dark states are mixed in the wall.}
\label{fig:bands}
\end{figure}

We consider an exciton in a two-dimensional semiconductor placed on, or next to, a magnetic layer hosting a skyrmion with magnetization direction $\nn(\mathbf r)$. Near a valley $\tau=\pm1$, the electron ($i=c$) and the hole ($i=v$) are described by
\begin{equation}
h_i=\frac{\mathbf p^2}{2m_i}+\mathbf b_i\cdot\bm\sigma,\qquad \mathbf b_i=J_i\,\nn(\mathbf r)+\eta_i\tau\Delta_i\,\hat{\mathbf z},
\label{eq:hi}
\end{equation}
where $J_i$ is the exchange coupling to the texture, $2\Delta_i$ the Ising spin--orbit splitting of the corresponding band and $\eta_i=\pm1$ its ordering. In transition-metal dichalcogenides, $\Delta_v$ ranges from several tens to a few hundred meV, while $\Delta_c$ lies between about 1.5 and 20\,meV (Fig.~\ref{fig:bands}a). The spin of the hole is therefore locked to the valley, whereas the spin of the electron can be reoriented by a modest exchange field. Bright excitons are formed by an electron and a hole of the same spin and couple to circularly polarized light, $\sigma^+$ in $K$ and $\sigma^-$ in $K'$; dark excitons involve an electron of the opposite spin. In semiconductors with weak spin--orbit coupling, both splittings are small.

The important observation is that the spin of a carrier does not follow the magnetization $\nn$ itself, but the total effective field $\mathbf b_i$, which also contains the Ising term. Far from the skyrmion, both $\nn$ and $\mathbf b_i$ point along $+\hat{\mathbf z}$. In the core, $\nn$ is reversed, but $\mathbf b_i$ is reversed only if $J_i>\Delta_i$. Otherwise the Ising field dominates and $\mathbf b_i$ keeps pointing up, so that its direction merely tilts in the skyrmion wall and returns. The emergent flux experienced by the carrier is given by the solid angle swept by $\hat{\mathbf b}_i$ (Fig.~\ref{fig:bands}b), which leads to
\begin{equation}
\frac{\Phi_i}{\Phi_0}=\frac{s_i\nu}{2}\big[\cos\Theta_i(0)-\cos\Theta_i(\infty)\big]=\begin{cases}\pm\nu,&J_i>\Delta_i,\\ 0,&J_i<\Delta_i,\end{cases}
\label{eq:fluxswitch}
\end{equation}
where $\nu$ is the topological charge of the skyrmion and $\Theta_i$ the polar angle of $\mathbf b_i$. As expected from topology, the flux is quantized. It is, however, either fully present or completely absent, and which of the two applies is decided by the competition between exchange and spin--orbit coupling. For an isolated carrier, the relevant quantity is the band splitting. In MoS$_2$, which has the smallest conduction-band splitting among the dichalcogenides, an exchange of a few meV is sufficient to switch on the flux for the electron, whereas the hole practically never winds with the texture. For an intralayer exciton, the relevant scale turns out to be the splitting between bright and dark excitons, which also contains the electron--hole exchange interaction, as we discuss in the section on materials.

We note that the sharp step in Eq.~\eqref{eq:fluxswitch} is a property of the adiabatic, projected description. Close to $J_i\approx\Delta_i$, the gap between the two spin branches closes in the skyrmion core, the carrier can no longer follow adiabatically, and the true crossover is smooth. The exact solution presented below shows that the emergent-flux picture becomes quantitative once $J_i$ exceeds both $\Delta_i$ and the kinetic energy scale $\hbar^2/(2m_iw^2)$ associated with a wall of width $w$ (Supplementary Note~\ref{sn:switch}).

To describe the exciton as a whole, we project each constituent on its adiabatic branch, introduce centre-of-mass and relative coordinates $\mathbf R$ and $\mathbf r$, and expand the emergent vector potentials to first order in $\mathbf r$ (Methods and Supplementary Note~\ref{sn:model}). This leads to the two-body Hamiltonian
\begin{equation}
H=\frac{\big[\mathbf P-\hbar\mathbf a_X-\hbar\Omega_-\hat{\mathbf z}\times\mathbf r\big]^2}{2M}+\frac{\big[\mathbf p-\tfrac{\hbar}{2}\Omega_+\hat{\mathbf z}\times\mathbf r\big]^2}{2\mu}+V_C(\mathbf r)+\sum_{i}(\varepsilon_i+W_i),
\label{eq:H2body}
\end{equation}
in which $\mathbf a_X=\mathbf a_e+\mathbf a_h$ is the emergent vector potential acting on the exciton, $\Omega_\pm$ are mass-weighted combinations of the emergent fields of the electron and the hole, $\varepsilon_i$ are the exchange energies and $W_i$ the geometric scalar potentials. For a real, uniform magnetic field, Eq.~\eqref{eq:H2body} reduces to the Hamiltonian of Gor'kov and Dzyaloshinskii~\cite{gorkov1968}, and $\mathbf a_X$ vanishes because the exciton is neutral. In a skyrmion this is no longer the case. The exciton carries no electric charge, but it does carry an emergent charge, and this charge depends on spin, valley and the ratio $J_i/\Delta_i$ of each constituent.

\begin{figure}[!tbp]
\centering\includegraphics[width=\textwidth]{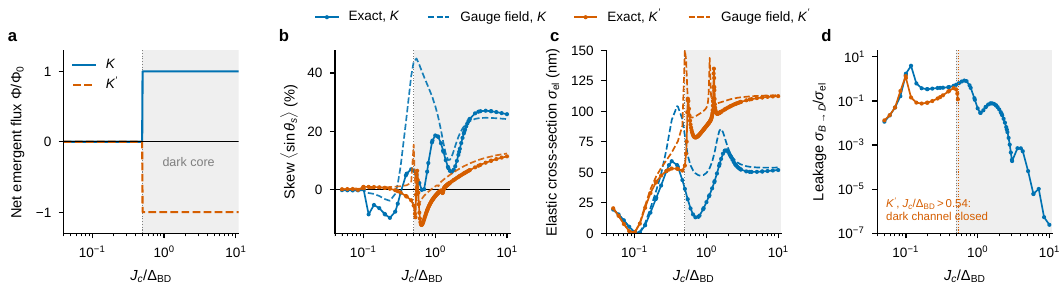}
\caption{\textbf{Spin--orbit coupling switches the emergent flux.} Skyrmion radius 10\,nm, wall parameter 4\,nm, $J_v=0.5$\,meV, bright--dark splitting $\Delta_\mathrm{BD}=5$\,meV, kinetic energy $k_\mathrm{B}T$ at 4.2\,K; $K$ valley in blue, $K'$ in orange. The shaded region, $J_c>J^*=\Delta_\mathrm{BD}/2$, marks the dark core. \textbf{a}, Net emergent flux of the branch connected to the bright exciton versus $J_c/\Delta_\mathrm{BD}$. \textbf{b}, Scattering skew and \textbf{c}, elastic cross-section from the exact two-channel solution (symbols) and from the gauge-field theory (dashed lines); the narrow gauge-field resonances in $K'$ are clipped at 150\,nm. \textbf{d}, Bright-to-dark leakage of the exact solution relative to elastic scattering, which signals the breakdown of adiabatic following. In $K'$, the dark channel is closed for $J_c/\Delta_\mathrm{BD}>0.54$ (dotted line), and there is no leakage.}
\label{fig:switch}
\end{figure}

The same physics can be formulated in the basis of bright and dark excitons, which differ by the spin of the electron. The in-plane component of the texture, $J_c\sin\theta\,e^{-i\tau\nu\varphi}$, couples the two states, and the phase winding of this coupling imposes a selection rule: a bright exciton with angular momentum $\ell$ can only be scattered into a dark exciton with angular momentum $\ell+\tau\nu$. In other words, each scattering event transfers one unit of angular momentum per unit of topological charge. The diagonal part of the coupling shifts the bright and dark excitons in opposite directions in the two valleys and reverses their valley ordering in the skyrmion core, while the off-diagonal part mixes them in the wall (Fig.~\ref{fig:bands}c). In the weak-coupling regime, a single skyrmion therefore appears as an island of reversed valley ordering surrounded by a ring of brightened dark excitons (Supplementary Fig.~\ref{fig:S4}). We have solved this two-channel scattering problem exactly and compared it with the gauge-field description of Eq.~\eqref{eq:H2body} (Fig.~\ref{fig:switch} and Supplementary Note~\ref{sn:weak}). Below the switch, the scattering is weak but chiral; its asymmetry reverses for an antiskyrmion and differs between the two valleys (Supplementary Fig.~\ref{fig:S3}). Close to the switch, the two descriptions disagree, and a substantial fraction of bright excitons is converted into dark ones. Well above the switch, this conversion drops below $10^{-3}$, and the skew of the scattering obtained from the exact solution (24--27\%) agrees with the gauge-field result (23--25\%). In $K'$, the bright exciton lies below the dark one once the flux is switched on, so that the dark channel is closed at thermal energies and no conversion occurs. The skew is smaller there and the cross-section about twice as large as in $K$.

This strong-coupling regime, in which every constituent winds with the texture, corresponds to the situation considered in an earlier study of excitons scattered by skyrmions in a CrI$_3$ monolayer, where the exciton was treated as a structureless spinor~\cite{kazemi2023}. The composite description contains this limit as a special case and adds the features that a structureless description cannot capture, namely the flux switch, the valley dependence and the crossover to weak coupling. We also note that for wave vectors accessible to optical excitation, $kR\ll1$, a gauge-consistent solution of the structureless model gives scattering asymmetries below $10^{-3}$ (Supplementary Note~\ref{sn:spinor} and Supplementary Fig.~\ref{fig:S2}). Sizeable chirality requires $kR\gtrsim1$, as in the calculations of Fig.~\ref{fig:switch}.

\subsection*{Trapping by quantum geometry}
We now turn to the fate of an exciton once the flux is switched on. Equation~\eqref{eq:H2body} contains two geometric contributions of different nature. The emergent vector potential $\mathbf a_X$ is the Berry connection of the texture, and its curl, the Berry curvature, integrates to the topological charge. The scalar potentials $W_i=\tfrac{\hbar^2}{8m_i}(\nabla\hat{\mathbf b}_i)^2$, on the other hand, are proportional to the quantum metric of the texture. They are largest where the spins rotate rapidly, that is, in the skyrmion wall. Both contributions enter the Hamiltonian as non-negative terms. On their own, they cannot bind an exciton, but they can hold it for a finite time behind a barrier.

\begin{figure}[!tbp]
\centering\includegraphics[width=\textwidth]{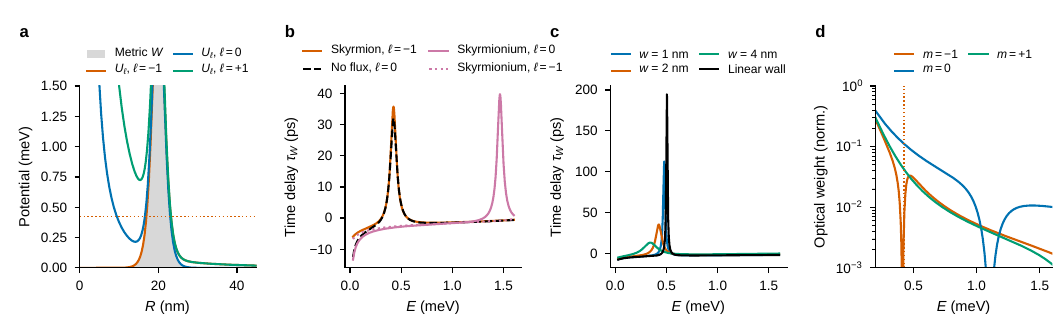}
\caption{\textbf{A trap made of quantum geometry} (radius 20\,nm, wall 2\,nm, $\nu=1$). \textbf{a}, The quantum-metric potential $W$ forms a ring-shaped barrier in the wall; the effective potentials $U_\ell$ include the centrifugal term shifted by the emergent flux. \textbf{b}, Wigner time delay: the resonance of the skyrmion in $\ell=-1$ survives when the emergent vector potential is removed, but reappears in $\ell=0$; a skyrmionium without net flux traps in $\ell=0$ as well. \textbf{c}, Dependence of the resonance on the wall width and shape. \textbf{d}, Optical weight seen by a 30\,nm near-field spot carrying orbital angular momentum $m$; the trapped $s$-mode (dotted line) appears as a feature only for $m=-\nu$.}
\label{fig:cavity}
\end{figure}

This is indeed what we find (Fig.~\ref{fig:cavity}). The quantum metric produces a ring-shaped barrier in the wall, and an exciton located in the core is trapped behind it. For a skyrmion with a radius of 20\,nm, the scattering phase shifts reveal a resonance at 0.43\,meV with a lifetime of about 9\,ps. To clarify the role of topology, we performed two control calculations (Fig.~\ref{fig:cavity}b). In the first, we removed the emergent vector potential and kept the metric. The resonance remains at the same energy, but it now appears in the channel $\ell=0$ instead of $\ell=-1$. In the second, we replaced the skyrmion by a skyrmionium, which has walls of the same type but no net topological charge. The skyrmionium also traps an exciton, again in $\ell=0$. We conclude that the barrier originates from the quantum metric, while the topology shifts the angular momentum under which the trapped state appears from $\ell$ to $\ell-\nu$. This shift is exact and fixed by the topological charge. The energy and the lifetime of the resonance, in contrast, depend on the details of the wall. They vary between 3 and 28\,ps when the wall width changes from 4 to 1\,nm and reach about 50\,ps for a linear wall profile (Fig.~\ref{fig:cavity}c). Within a given family of profiles, the lifetime increases with the topological charge and with the size of the skyrmion, from picoseconds to hundreds of picoseconds (Supplementary Note~\ref{sn:trap} and Supplementary Fig.~\ref{fig:S7}). These are useful trends, but they are not protected by topology.

If the exchange energies $\varepsilon_i$ add an attractive well, the resonances become true bound states and the same rule applies. The ground state carries the angular momentum $\ell=-\nu$, and the excited $p$-like states, which would be degenerate in an ordinary circular trap, are split by the emergent field (Supplementary Fig.~\ref{fig:S7}).

The angular-momentum label has a direct optical consequence. Light couples to the bright component of the trapped exciton, and light carrying an orbital angular momentum $m$ couples only to states with $\ell=m$. We have therefore computed (Supplementary Note~\ref{sn:trap}) the optical weight seen by a near-field spot of 30\,nm diameter carrying orbital angular momentum $m$ (Fig.~\ref{fig:cavity}d). The $s$-like trapped mode appears as a feature at 0.43\,meV only for $m=-\nu$. Light without orbital angular momentum addresses a different trapped mode near 1\,meV, and light of the opposite handedness addresses neither. Reversing the topological charge of the skyrmion interchanges the roles of $m=+1$ and $m=-1$. The corresponding signals are weak, at most a few per cent of the optical weight at the band edge, because the trapped modes are located in the core, where the exciton is spin-dark, and because they lie at momenta outside the light cone. They nevertheless constitute a direct optical fingerprint of the topological charge.

\subsection*{A trap that moves}
An exciton bound in a quantum dot stays where the dot is. A skyrmion, in contrast, can be displaced along a track by a current, and the natural question is whether a bound exciton follows it.

\begin{figure}[!tbp]
\centering\includegraphics[width=\textwidth]{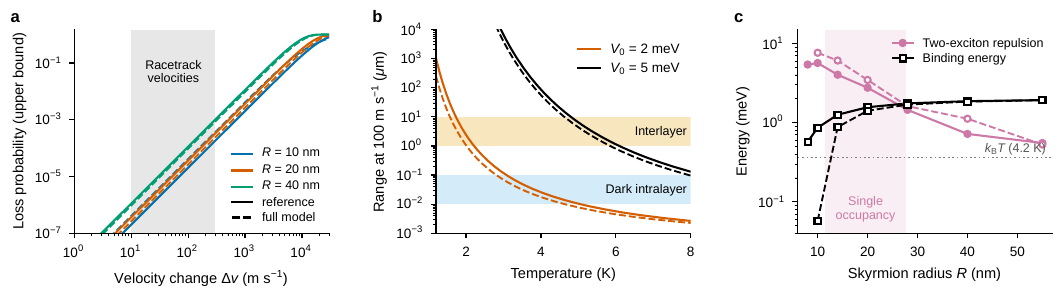}
\caption{\textbf{The skyrmion as a mobile carrier of excitons.} Interlayer excitons in MoS$_2$/WSe$_2$ ($M=0.83\,m_0$) bound in the exchange well of a skyrmion with wall width 1.5\,nm. The well depth is $V_0=2$\,meV unless stated otherwise. Solid lines: reference model with the exchange well only; dashed lines: full model including the Berry and geometric terms (Methods). \textbf{a}, Upper bound for the probability of losing the exciton when the skyrmion velocity changes instantaneously by $\Delta v$, for skyrmion radii of 10, 20 and 40\,nm. The grey band marks typical velocities of current-driven skyrmions. \textbf{b}, Distance $v\tau_{\rm esc}$ over which the exciton is carried at $v=100$\,m/s before it escapes thermally, as a function of temperature, for a skyrmion radius of 20\,nm and well depths of 2 and 5\,meV, which give binding energies of 1.6 and 4.5\,meV. The coloured bands show the distances $v\tau_X$ allowed by the lifetimes $\tau_X$ of dark intralayer (0.1--1\,ns) and interlayer (10--100\,ns) excitons. The achievable range is the smaller of the two. \textbf{c}, Dipolar repulsion between two interlayer excitons occupying the ground state of the same skyrmion (layer separation 0.65\,nm, dielectric constant 4.5), compared with the binding energy of a single exciton. The dotted line marks $k_BT$ at 4.2\,K. In the shaded region, the full model gives a binding energy above $k_BT$ and a repulsion above the binding energy, so that the skyrmion holds exactly one exciton.}
%\caption{\textbf{The skyrmion as a mobile carrier of excitons} (interlayer exciton, $M=0.83\,m_0$, exchange well of 2\,meV, wall width 1.5\,nm). \textbf{a}, Upper bound for the probability of losing the exciton after an instantaneous velocity change $\Delta v$; the grey band marks racetrack velocities. \textbf{b}, Transport range at 100\,m/s limited by thermal escape for two binding energies, compared with the ranges set by the lifetimes of dark intralayer and interlayer excitons. \textbf{c}, Dipolar energy of two interlayer excitons in the same skyrmion compared with the binding energy; below a radius of about 25\,nm the skyrmion holds a single exciton.}
\label{fig:carrier}
\end{figure}

For transport, the metastable resonances of the purely geometric trap are not sufficient, and we consider the bound states that form when the exchange energy of the constituents adds an attractive well of depth $V_0$ in the skyrmion core, as discussed above. We treat $V_0$, which is of the order of the exchange coupling, as a parameter and use 2 and 5\,meV. As a simple reference, we first omit the Berry and geometric terms and keep only the exchange well; we then compare with the full model, which includes both (solid and dashed lines in Fig.~\ref{fig:carrier}).

A simple and exact argument answers the first part of this question. Since the problem is invariant under Galilean transformations, a skyrmion moving at constant velocity carries its bound exciton without any loss. Losses can only arise while the skyrmion accelerates. The most unfavourable case is an instantaneous change of velocity $\Delta v$, for which the probability of retaining the exciton is $|\langle\psi_0|e^{iM\Delta v\cdot x/\hbar}|\psi_0\rangle|^2$. We evaluated this bound for interlayer excitons, which turn out to be the most suitable candidates, as discussed below (Supplementary Note~\ref{sn:transport}). Even an instantaneous jump to 100\,m/s removes at most about $10^{-4}$ of the exciton (Fig.~\ref{fig:carrier}a), and losses become significant only above about 10\,km/s, far beyond any racetrack velocity. Similarly, tunnelling out of the trap during acceleration would require accelerations of the order of $10^{17}$\,m/s$^2$, whereas a skyrmion that reaches 100\,m/s~\cite{woo2016} within the 0.1--10\,ns
rise time of a current pulse experiences only
$a\approx\Delta v/\tau\approx10^{10}$--$10^{12}$\,m/s$^2$.
The same holds in the full model. With the Berry potential, the moving texture also produces an emergent electric field, but in the frame moving with the skyrmion this only adds the static term $-\hbar\mathbf v\cdot\mathbf a$, of a few $\mu$eV at 100\,m/s, and the losses remain of the same order (Fig.~\ref{fig:carrier}a). For a rigid skyrmion in a clean system, the motion itself is therefore not a limitation. Disorder, deformations of the skyrmion during its motion and exciton--phonon scattering are not included in this analysis and will reduce the transport distance.

What does limit transport is time. The distance over which an exciton can be carried is the skyrmion velocity multiplied by the shorter of two times, the lifetime of the exciton and the time it needs to escape thermally from the trap. Figure~\ref{fig:carrier}b compares the two. The curves show the distance covered before thermal escape, and the bands the distance covered within the exciton lifetime; whichever is shorter sets the range. Bright intralayer excitons live only for picoseconds, and even dark intralayer excitons for at most about a nanosecond. Interlayer excitons in heterobilayers of transition-metal dichalcogenides, in which the electron and the hole reside in different layers, live for tens of nanoseconds or longer~\cite{Miller_2017}. They are attractive for a second reason. If the layer with the smaller Ising splitting faces the magnet, only the electron couples to the texture and carries the full emergent charge, and the flux switch then requires no more than a few meV of exchange. For a binding energy of 1.6\,meV, thermal escape limits the range to a few micrometres at 2\,K and to well below a micrometre above about 3\,K. For a binding energy of 4.5\,meV, thermal escape becomes slower than the lifetime of interlayer excitons below roughly 5\,K, and the exciton can be carried over several micrometres at 100\,m/s, limited by its lifetime. At 4.2\,K, the escape-limited distance is then about 50\,$\mu$m. Including the Berry and geometric terms reduces the binding energy at this radius by 10\% for $V_0=2$\,meV and by 2\% for $V_0=5$\,meV, which shortens the escape time by less than a factor of two, well within the uncertainty of the prefactor. These estimates inherit the uncertainty of the escape prefactor, which we estimate to be about one order of magnitude, but the resulting design rule is robust: the binding energy should exceed about ten times $k_BT$.

Interlayer excitons have one further property that matters here. They carry a permanent out-of-plane electric dipole and therefore repel each other. 

We evaluated this repulsion for two excitons occupying the ground state of the same skyrmion. For skyrmions with radii below about 25--27\,nm, it exceeds the binding energy of a single exciton (Fig.~\ref{fig:carrier}c), and such a skyrmion accepts one exciton but not a second one. Here the full model adds an important lower bound. In small skyrmions, the geometric barrier in the wall pushes the bound state up, and for a well of 2\,meV the binding energy falls below $k_BT$ at 4.2\,K for radii below about 12\,nm. A single exciton is therefore held stably in skyrmions with radii between about 12 and 27\,nm. Deeper wells extend this window to smaller radii; for $V_0=5$\,meV the binding energy at a radius of 10\,nm is still 2.8\,meV. The precise crossover radius depends on the dielectric environment and on the layer separation, and it should be regarded as an estimate to be refined for a specific heterostructure. Within these limits, a small skyrmion becomes a mobile carrier of single excitons, which can be loaded optically, moved by a current pulse and emptied at a chosen position. In combination with skyrmion logic operations such as merging and duplication~\cite{zhang2015logic}, this blockade could provide the nonlinearity needed to process excitonic information.

\subsection*{Candidate materials}
The map in Fig.~\ref{fig:map} summarizes the conditions identified so far. First, the flux has to be switched on, which requires an exchange coupling above the relevant spin--orbit splitting. Second, the exciton has to follow the texture adiabatically, which favours wide skyrmion walls. Third, trapping and transport require a binding energy above $k_BT$, and preferably above $10\,k_BT$, which favours large skyrmions. Two-dimensional excitonic materials fall into a few families with respect to these criteria (Fig.~\ref{fig:map}c, Supplementary Note~\ref{sn:materials} and Supplementary Tables~\ref{tab:S1} and \ref{tab:S2}).

The transition-metal dichalcogenides are the best characterized excitonic materials, and in their case a subtlety has to be taken into account. For an intralayer exciton, the texture mixes bright and dark states, and the flux switch is therefore controlled by the bright--dark splitting $\Delta_{BD}$, with a threshold $J^*=|\Delta_{BD}|/2$, rather than by the bare splitting of the conduction band. The two quantities differ because $\Delta_{BD}$ also contains the short-range electron--hole exchange and the different binding energies of bright and dark excitons. Magneto-optical measurements on hBN-encapsulated monolayers give $\Delta_{BD}=+14$\,meV for MoS$_2$, with the dark exciton below the bright one, and $\Delta_{BD}=-1.4$\,meV for MoSe$_2$, with the dark exciton slightly above, even though the bare conduction-band splittings are 3 and 22\,meV, respectively~\cite{robert2020,kormanyos2015}. For intralayer excitons, MoSe$_2$ is therefore the most favourable dichalcogenide, with a threshold below 1\,meV. MoS$_2$ requires about 7\,meV, and the W-based compounds, in which the dark excitons lie 40--55\,meV below the bright ones, 20\,meV or more~\cite{robert2020}.

Our $GW$–Bethe–Salpeter equation (BSE) calculations for free-standing MoS$_2$ support this picture. They give a bright A exciton at 1.91\,eV, bound by 0.82\,eV with respect to the quasiparticle gap, and place the spin-forbidden dark A exciton 23\,meV below it (Supplementary Note~\ref{sn:gw} and Supplementary Fig.~\ref{fig:S1}). This is the same ordering as observed in hBN-encapsulated samples, where the additional screening reduces the splitting to 14\,meV~\cite{robert2020}. The same experiments also provide a useful calibration of the coupling between bright and dark states that underlies our results. An in-plane magnetic field brightens the dark exciton of MoS$_2$, which becomes visible above about 12\,T~\cite{robert2020}. With an in-plane electron $g$-factor of 2, an exchange coupling of 1\,meV acting on the electron is equivalent to an in-plane field of about 17\,T. A skyrmion wall with this exchange therefore acts on the exciton like a local in-plane field in the range where the brightened dark exciton has already been observed.

Heterobilayers offer an additional degree of freedom, because the electron and the hole of an interlayer exciton reside in different layers and only the layer in contact with the magnet matters. If MoS$_2$ faces the magnet and WSe$_2$ hosts the hole, only the electron couples to the texture. Since the electron--hole exchange is weak for spatially separated carriers, the threshold is then close to the bare conduction-band value of MoS$_2$, about 1.5\,meV. In MoSe$_2$/WSe$_2$, the most thoroughly characterized heterobilayer, the electron resides in MoSe$_2$ and the threshold increases to about 10\,meV, so that the most studied system is not the most favourable one in this context. Interlayer excitons are larger than intralayer ones, live for nanoseconds or longer~\cite{rivera2015,rivera2016} and diffuse over micrometres~\cite{unuchek2018,jauregui2019}, which makes them well suited as cargo. Two aspects require care. In aligned heterobilayers, moir\'e potentials of tens of meV trap interlayer excitons by themselves~\cite{seyler2019}. At large twist angles, however, the moir\'e period of a few nanometres is averaged out over a skyrmion of tens of nanometres. In addition, the optical selection rules of interlayer excitons depend on the stacking registry, and spin-triplet configurations can be optically active. In that case, the spin-dark core of the intralayer case would become a core emitting light of the opposite circular polarization, which would be an equally useful signature.

Other families are located in different parts of the map. Semiconductors with weak spin--orbit coupling, such as black phosphorus or InSe, have no Ising splitting to overcome, so that any exchange coupling winds the carrier spins, but they lack the valley selectivity of the dichalcogenides. In magnetic semiconductors such as CrSBr, or CrI$_3$, in which skyrmions have been proposed~\cite{liu2018,behera2019}, the carriers are themselves magnetic, and these materials sit naturally in the strong-coupling corner considered in ref.~\cite{kazemi2023}. Antiferromagnetic excitonic materials such as NiPS$_3$ and MnPS$_3$ form a special case, since the fluxes from the two sublattices cancel and only valley- or spin-resolved effects are expected to survive.

A further question is whether skyrmion hosts and excitonic layers can be combined without destroying the skyrmions. As a test case, we considered Pd/Fe/Ir(111), in which individual skyrmions can be written and deleted on demand~\cite{romming2013,romming2015}, covered by two hBN layers and a MoS$_2$ monolayer. Our first-principles calculations, performed using the fully relativistic Korringa–Kohn–Rostoker (KKR) Green-function method~\cite{russmann2022judftteam}, show that the isotropic exchange and the Dzyaloshinskii--Moriya interaction that stabilize the skyrmions are essentially unaffected by the van der Waals overlayer (Supplementary Note~\ref{sn:pdfe} and Supplementary Fig.~\ref{fig:S5}). Pd/Fe/Ir(111) also illustrates the small-size limit of the map, with skyrmion diameters of only 2--4\,nm. Even in this limit, the valley-split island in the core and the ring of brightened dark excitons in the wall retain most of their contrast after averaging over the extent of the exciton (Supplementary Fig.~\ref{fig:S6}). For the excitonic coupling itself, however, an insulating skyrmion host in direct contact with the excitonic layer is preferable to a metallic one, which requires a spacer to protect the exciton.

These considerations suggest a concrete experiment. A MoSe$_2$ monolayer, or a MoS$_2$/WSe$_2$ heterobilayer for transport, is placed in direct contact with an insulating magnet that hosts skyrmions, for instance the chiral magnet Cu$_2$OSeO$_3$, in which skyrmions were first observed in an insulator~\cite{seki2012}. Polarization-resolved near-field photoluminescence could then image the valley-split island and the ring of brightened dark excitons, the latter being the local counterpart of the brightening observed in uniform in-plane fields~\cite{robert2020}. Writing and erasing individual skyrmions~\cite{romming2013} would switch these signatures on and off, and the trapped states would appear as emission lines whose energy follows the size of the skyrmion as the magnetic field is varied.

\section*{Discussion}
The picture that emerges from our analysis differs in several respects from what one would expect by analogy with electrons. Because the exciton is a composite particle, spin--orbit coupling decides whether it experiences the topology of a skyrmion at all, and the two valleys experience it with opposite signs. When the flux is present, the quantum geometry of the texture traps the exciton. The quantum metric provides the barrier, while the topology fixes the angular momentum of the trapped states and, through it, the light that can address them. In contrast to a conventional quantum dot, this trap can be moved by a current together with its exciton. Throughout this work, we have tried to distinguish clearly between the properties fixed by topology, namely the flux quantum and the shift of angular momentum, and those that depend on details of the system, such as energies, lifetimes and transport distances.

Several extensions appear natural. The quantitative values of the proximity exchange at interfaces between excitonic semiconductors and insulating skyrmion hosts remain to be determined, and first-principles calculations for specific heterostructures would sharpen the predictions of Fig.~\ref{fig:map}. Disorder, deformations of moving skyrmions and exciton--phonon scattering, which we have not included, will set the practical limits of transport. Finally, the results are not restricted to excitons. Any neutral composite particle in a texture-induced gauge field, including cold atoms in synthetic gauge potentials~\cite{dalibard2011}, should display the same flux switch and the same type of geometric trap.

Excitons and skyrmions have so far been studied largely in separate communities, one optical and one magnetic. The possibility of trapping excitons in skyrmions, labelling them by topology and moving them with currents could motivate experiments that bring the two together. This defines an exciting research road for the realization of hybrid devices for future information technology, where both worlds meet: topological magnetism and optical excitations.

\section*{Methods}
The essential steps of the theory and the equations used in the calculations are summarized here. The full derivations are given in Supplementary Notes~\ref{sn:model}, \ref{sn:switch}, \ref{sn:weak}, \ref{sn:trap} and \ref{sn:transport}, and Supplementary Note~\ref{sn:curves} lists, for every curve shown, the equation that was evaluated and the parameters that were used.

\paragraph{Excitons from first principles.} Ground-state calculations of monolayer MoS$_2$ were performed with \textsc{Quantum Espresso}~\cite{qe,giannozzi2009} using Optimized Norm-Conserving Vanderbilt (ONCV) pseudopotentials~\cite{hamann2013}, the Perdew-Burke-Ernzerhof (PBE) functional~\cite{perdew1996} and spin--orbit coupling. Quasiparticle energies were obtained within $G_0W_0$~\cite{Hedin1965, Hybertsen1985,Rohlfing2000,Onida2002} with the generalized plasmon-pole model~\cite{godby1989} and a truncated Coulomb interaction~\cite{rozzi2006}, and the BSE~\cite{Rohlfing2000,Onida2002,Strinati1988} was solved with \textsc{Yambo}~\cite{marini2009} on a $27\times27\times1$ $k$-mesh{\cite{Monkhorst1976}}, including spin--orbit coupling, so that bright and spin-forbidden dark excitons are obtained on the same footing. At $K$ we find spin--orbit splittings of 149.5\,meV (valence) and 1.6\,meV (conduction) in $GW$. For the model, we use the parameters of hBN-encapsulated monolayers measured by high-field magneto-optics~\cite{goryca2019}: for MoS$_2$, a reduced mass $\mu=0.275\,m_0$, a binding energy of 221\,meV and an rms radius of the 1s exciton of 1.2\,nm, consistent with the extent of the BSE density in Supplementary Fig.~\ref{fig:S1}. Band masses are taken from $\mathbf{k}\cdot\mathbf{p}$ fits to first-principles bands~\cite{kormanyos2015} ($m_e\approx0.44$--$0.47\,m_0$, $m_h\approx0.54\,m_0$ for MoS$_2$, $M\approx1.0\,m_0$).

The magneto-optical reduced mass, $\mu=0.275\,m_0$, points to a heavier electron
($m_e\approx0.56\,m_0$ for $m_h=0.54\,m_0$)~\cite{goryca2019}, i.e.,
$M\approx1.1\,m_0$; heavier excitons bind more strongly in a given trap, so that
our estimates of binding and transport are conservative.
In the model calculations we use a representative bright--dark splitting $\Delta_{BD}=5$\,meV, between the measured values of MoSe$_2$ ($-1.4$\,meV) and MoS$_2$ ($+14$\,meV)~\cite{robert2020}; the dark-exciton ring weight scales as $(J_c/\Delta_{BD})^2$, so it is weaker in MoS$_2$ and stronger in MoSe$_2$.

\paragraph{Magnetic interactions from first principles.} We modelled
Pd/Fe/Ir(111), without and with an overlayer of two hBN layers and a MoS$_2$
monolayer, by a slab of five Ir(111) layers covered by Fe, Pd and the overlayer,
with a $2\times2$ in-plane supercell ($a=2.715$\,\AA) and about 24\,\AA{} of
vacuum. The structure was relaxed with \textsc{Quantum Espresso}~\cite{qe} and PAW
pseudopotentials~\cite{ps} (plane-wave cutoff 90\,Ry, $12\times12\times1$
$k$-points), optimizing the atomic positions and the in-plane lattice vectors until
the forces were below $10^{-5}$\,Ry/$a_0$. The magnetic interactions were then
computed with the full-potential Korringa--Kohn--Rostoker Green-function method
(JuKKR~\cite{russmann2022judftteam}) in the local density
approximation~\cite{vosko1980}, with spin--orbit coupling included
self-consistently, $\ell_{\max}=3$ and an electronic temperature of 502.78\,K on
the complex energy contour. The isotropic exchange $J_{ij}$ and the
Dzyaloshinskii--Moriya vectors $\mathbf D_{ij}$~\cite{DZYALOSHINSKY1958241}
were obtained from infinitesimal rotations in the ferromagnetic
state~\cite{LIECHTENSTEIN198765,PhysRevB.79.045209} on a $200\times200\times1$
$k$-mesh, for all Fe pairs up to $8a$.

\paragraph{Skyrmion profiles.} We describe a N\'eel skyrmion of charge $\nu$ by $\nn=(\sin\theta\cos\nu\varphi,\sin\theta\sin\nu\varphi,\cos\theta)$ with the 360$^\circ$-wall profile $\theta(R)=\pi+\sum_\pm\arcsin\tanh[(R\pm c)/w]$, which describes measured skyrmions accurately~\cite{romming2015}; $c$ is the radius and $w$ the wall parameter. For Pd/Fe/Ir(111), $c$ and $w$ were obtained by minimizing the micromagnetic energy of ref.~\cite{romming2015} within this profile (Supplementary Note~\ref{sn:pdfe}).

\paragraph{Emergent fields of a single carrier.} The effective field $\mathbf b_i$ of Eq.~\eqref{eq:hi} lies in the vertical plane containing $\nn$. Its azimuth is $\nu\varphi$ and its polar angle $\Theta_i$ follows from $\cos\Theta_i=(J_i\cos\theta+\eta_i\tau\Delta_i)/|\mathbf b_i|$. Projecting $h_i$ onto the spinor $\chi_i$ of the branch $s_i=\pm1$ gives
\begin{equation}
h_i^{\rm ad}=\frac{(\mathbf p-\hbar\mathbf a_i)^2}{2m_i}+s_i|\mathbf b_i|+W_i,\qquad
\mathbf a_i=-\frac{1-\cos\Theta_i}{2}\,\nabla\phi_i,\qquad
W_i=\frac{\hbar^2}{8m_i}\big[(\nabla\Theta_i)^2+\sin^2\Theta_i(\nabla\phi_i)^2\big],
\end{equation}
where $\mathbf a_i=i\braket{\chi_i|\nabla\chi_i}$ is the Berry connection and $W_i$ is $\hbar^2/2m_i$ times the trace of the quantum metric of $\chi_i$. For the hole, $\mathbf a_h=-\mathbf a_v$. The flux of $\nabla\times\mathbf a_i$ is the solid angle swept by $\hat{\mathbf b}_i$, which gives Eq.~\eqref{eq:fluxswitch}. The $z$ component of $\mathbf b_i$ equals $J_i+\eta_i\tau\Delta_i$ outside the skyrmion and $-J_i+\eta_i\tau\Delta_i$ in the core; the two have opposite signs, and the flux is a full quantum, only if $J_i>\Delta_i$.

\paragraph{Two-body Hamiltonian.} With $\mathbf R=\alpha\mathbf r_e+\beta\mathbf r_h$, $\mathbf r=\mathbf r_e-\mathbf r_h$, $\alpha=m_e/M$, $\beta=m_h/M$ and $\mu=m_em_h/M$, we expand $\mathbf a_e(\mathbf R+\beta\mathbf r)$ and $\mathbf a_h(\mathbf R-\alpha\mathbf r)$ to first order in $\mathbf r$. The symmetric part of $(\mathbf r\cdot\nabla)\mathbf a$ is a gradient in $\mathbf r$ and is gauged away, leaving $\tfrac12\Omega\,\hat{\mathbf z}\times\mathbf r$ with $\Omega=\hat{\mathbf z}\cdot(\nabla\times\mathbf a)$. The kinetic energy then separates exactly into $\bm\Pi^2/2M+\bm\pi^2/2\mu$, with $\bm\Pi=\mathbf P-\hbar\mathbf a_X-\tfrac{\hbar}{2}\Omega_-\hat{\mathbf z}\times\mathbf r$ and $\bm\pi=\mathbf p-\hbar(\beta\mathbf a_e-\alpha\mathbf a_h)-\tfrac{\hbar}{2}\Omega_+\hat{\mathbf z}\times\mathbf r$. The $\mathbf r$-independent term of $\bm\pi$ is removed by $U=\exp[i\mathbf r\cdot(\beta\mathbf a_e-\alpha\mathbf a_h)]$. Because $U$ depends on $\mathbf R$, it also shifts $\mathbf P$ by $\hbar\nabla_{\mathbf R}[\mathbf r\cdot(\beta\mathbf a_e-\alpha\mathbf a_h)]$, whose antisymmetric part is $-\tfrac{\hbar}{2}\Omega_-\hat{\mathbf z}\times\mathbf r$. This doubles the corresponding term in $\bm\Pi$ and gives Eq.~\eqref{eq:H2body}. Averaging over the 1s state, terms linear in $\mathbf r$ and the orbital Zeeman term vanish, and the emergent diamagnetic shift $\hbar^2\langle r^2\rangle[\Omega_-^2/2M+\Omega_+^2/8\mu]$ remains. In a uniform field, $\hbar\mathbf a_{e,h}=\mp e\mathbf A$, and Eq.~\eqref{eq:H2body} reduces to the Gor'kov--Dzyaloshinskii Hamiltonian~\cite{gorkov1968}.

\paragraph{Bright--dark Hamiltonian and threshold.} With the hole spin pinned, the texture acts on the bright ($B$) and dark ($D$) excitons through the electron spin. For an exciton small compared with the skyrmion, and measuring energies from the bright exciton,
\begin{equation}
H_\tau(\mathbf R)=\begin{pmatrix}\tau(J_c-J_v)\cos\theta & J_c\sin\theta\,e^{-i\tau\nu\varphi}\\ J_c\sin\theta\,e^{i\tau\nu\varphi} & -\Delta_\mathrm{BD}-\tau(J_c+J_v)\cos\theta\end{pmatrix}.
\label{eq:Htau}
\end{equation}
Written as $d_0+\mathbf b\cdot\bm\sigma$, the pseudospin field has $b_x-ib_y=J_c\sin\theta\,e^{-i\tau\nu\varphi}$ and $b_z=\tau J_c\cos\theta+\Delta_\mathrm{BD}/2$. This is the single-carrier problem with exchange $J_c$ and Ising term $\Delta_\mathrm{BD}/2$, so the branch connected to the bright exciton acquires a flux quantum when $J_c>J^*=|\Delta_\mathrm{BD}|/2$, the threshold used in Fig.~\ref{fig:map}c. The phase $e^{i\tau\nu\varphi}$ couples a bright partial wave $\ell$ only to a dark wave $\ell+\tau\nu$. For skyrmions comparable to the exciton, $\nn$ is replaced by its average over the 1s density as seen by the electron at $\mathbf R+\beta\mathbf r$ and by the hole at $\mathbf R-\alpha\mathbf r$ (Supplementary Note~\ref{sn:weak}).

\paragraph{Exact and adiabatic scattering.} For each $\ell$, the radial functions $u_B$ (angular momentum $\ell$) and $u_D$ ($\ell+\tau\nu$) obey
\begin{equation}
u_c''=\Big[\frac{m_c^2-\frac14}{R^2}+\frac{E_c(R)-E}{t_M}\Big]u_c+\frac{J_c\sin\theta}{t_M}\,u_{\bar c},\qquad t_M=\frac{\hbar^2}{2M},
\end{equation}
with $E_c(R)$ the diagonal elements of Eq.~\eqref{eq:Htau} and $E=\tau(J_c-J_v)+\varepsilon$. Two regular solutions are integrated outward (DOP853) and matched to $\sqrt R\,H^{(2)}_{m_c}(k_cR)$ (amplitude $A_c$) and $\sqrt R\,H^{(1)}_{m_c}(k_cR)$ (amplitude $B_c$). Since the outgoing flux is proportional to $|B_c|^2$ independently of $k_c$, $S=BA^{-1}$ is flux-normalized; it is unitary to better than $10^{-3}$. The cross-sections are $\sigma_{\rm el}=k_B^{-1}\sum_\ell|S^\ell_{BB}-1|^2$ and $\sigma_{B\to D}=k_B^{-1}\sum_\ell|S^\ell_{DB}|^2$, and the skew is $\langle\sin\theta_s\rangle=\int\sin\theta_s|f_{BB}|^2/\int|f_{BB}|^2$ with $f_{BB}=(2\pi k_B)^{-1/2}\sum_\ell(S^\ell_{BB}-1)e^{i\ell\theta_s}$. In the adiabatic limit, we follow the eigenvector $v=(v_B,v_D)$ of Eq.~\eqref{eq:Htau} that is bright at infinity and solve one radial equation with the Berry potential $A(R)=-\tau\nu[|v_D(R)|^2-|v_D(\infty)|^2]$ (gauge regular at infinity) and the geometric potential $W=t_M[|\partial_Rv|^2+\nu^2|v_B|^2|v_D|^2/R^2]$. The net flux is $\tau\nu|v_D(0)|^2$.

\paragraph{Resonances, bound states and optical weights.} For an exciton that follows the texture completely, the effective radial potential is
\begin{equation}
U_\ell(R)=t_M\frac{(\ell-A)^2-\frac14}{R^2}+W+V,\qquad A=\frac{\nu}{2}(\cos\theta-1),\qquad W=\frac{t_M}{4}\Big[\theta'^2+\nu^2\frac{\sin^2\theta}{R^2}\Big],
\end{equation}
with an optional exchange well $V=-V_0(1-\cos\theta)/2$. Phase shifts $\delta_\ell$ are obtained by matching to $\sqrt R[aJ_{|\ell|}(kR)+bY_{|\ell|}(kR)]$, $\tan\delta_\ell=-b/a$, and resonances from the Wigner time delay $\tau_W=2\hbar\,d\delta_\ell/dE$, with $\Gamma=4\hbar/\tau_W^{\max}$ and lifetime $\hbar/\Gamma$. The controls use $A\equiv0$ (no flux) and the skyrmionium profile $\theta=\tfrac{\pi}{2}[2+\tanh\tfrac{R-c_1}{w}+\tanh\tfrac{R-c_2}{w}]$. Bound states are obtained on a cell-centred finite-volume grid ($R_{\max}=300$--$400$\,nm). The optical weight for a near-field spot with orbital angular momentum $m$ is $S_m(E)=k^{-1}[\int dR\,\sqrt R\,g_m\,v_B\,u_E]^2$, with $g_m=(R/s)^{|m|}e^{-R^2/2s^2}$, $s=30$\,nm, the bright amplitude $v_B^2=(1+\cos\theta)/2$ and $u_E$ normalized to unit asymptotic amplitude.

\paragraph{Transport.} For a trap moving at constant velocity $\mathbf v$, $\psi=\exp[\tfrac{i}{\hbar}(M\mathbf v\cdot\mathbf R-\tfrac12Mv^2t-E_0t)]\psi_0(\mathbf R-\mathbf vt)$ solves the time-dependent problem exactly, so no loss occurs. After an instantaneous velocity change $\Delta v$, the probability of remaining bound is
\begin{equation}
P(\Delta v)=\Big|\int_0^\infty2\pi R\,dR\,|\psi_0(R)|^2J_0(qR)\Big|^2,\qquad q=\frac{M\Delta v}{\hbar},
\end{equation}
an upper bound for the loss at any finite acceleration. In the full model, the ground state carries $\ell=-\nu$, but its density is still rotationally symmetric, so the same expression applies. In the reference model of Fig.~\ref{fig:carrier}, $A$ and $W$ are set to zero. Thermal escape is estimated as $\tau_{\rm esc}=\nu_0^{-1}e^{E_b/k_BT}$ with $\nu_0=(E_1-E_0)/h$ from the two lowest $\ell=0$ levels; this prefactor is uncertain by about an order of magnitude. The interaction of two interlayer excitons in the trap is evaluated in the Hartree approximation, $U_{\rm dip}=\langle\frac{e^2}{4\pi\varepsilon_0\varepsilon}(\frac{2}{s}-\frac{2}{\sqrt{s^2+d^2}})\rangle$, by Monte Carlo sampling of the ground-state density, with $d=0.65$\,nm and $\varepsilon=4.5$. For interlayer excitons in MoS$_2$/WSe$_2$ we use $M=0.83\,m_0$, the sum of the first-principles electron mass of MoS$_2$ and hole mass of WSe$_2$~\cite{kormanyos2015} ($m_e\approx0.44$--$0.47$, $m_h\approx0.36\,m_0$); with the heavier electron mass inferred from magneto-optics~\cite{goryca2019}, $M\approx1.06\,m_0$, which again makes our estimates conservative (Supplementary Table~\ref{tab:S2}).

\section*{Data availability}
The data that support the findings of this study are present in the paper and the Supplementary Information.
%The data that support the plots and findings of this paper are included in theavailable from the authors.
%Samir Lounis \\\url{(samir.lounis@physik.uni-halle.de)} 
%upon reasonable request. \todo{Check: Nature Communications expects Source Data for all graphs and generally discourages ``upon request''; refer to the Source Data file. A Code availability statement is also required.}

\section*{Code availability}

The first-principles codes \textsc{Quantum Espresso}~\cite{giannozzi2009} and \textsc{Yambo}~\cite{marini2009} are freely available. The Korringa--Kohn--Rostoker Green-function code JuKKR, developed in J\"ulich, is also available~\cite{russmann2022judftteam}; It is a rather complex ab-initio DFT-based code, which is in general impossible to use without proper training on the theory behind it and on the practical utilization of the code. The Python codes used for the model calculations and
for all figures, together with the input files needed to reproduce
the figures, are available at \url{https://doi.org/10.5281/zenodo.24011147.}

%The ab-initio codes are freely available. The in-house developed KKR code [Refs] is also available. It is a rather complex ab-initio DFT-based code, which is in general impossible to use without proper training on the theory behind it and on the practical utilization of the code. The version that we used is the Jülich-developed codes JuKKR can be found at ~\cite{russmann2022judftteam}.
%The Python codes used for the model calculations and
%for all figures, together with the input files needed to reproduce
%the figures, are available at \selcuk{[repository, DOI]}

\section*{Acknowledgements}
Funding has been provided by the Priority Programme SPP 2137 ``Skyrmionics'' (Project LO 1659/8-1) of the Deutsche Forschungsgemeinschaft (DFG). We also acknowledge funding through
Project No.~328545488 -- CRC/TRR 227, Project No.~B12, and by the German Excellence Strategy -- EXC 3112/1 -- 533767171 (Center for Chiral Electronics). K.R. acknowledges funding from the Open Research Fund of the General Direction of Research and Technological Development (DGRSDT) of the Ministry of Higher Education and Scientiﬁc Research, Algeria.  The authors  disclose the use of AI tools to assist with drafting and refining portions of the manuscript based on author-provided scientific ideas and context that were critically reviewed, verified, and revised.

\section*{Author contributions}
S.L. initiated, designed, and supervised the project. K.R. developed the theoretical framework with the support of S.Z. and supervision of S.L. Simulations were performed by S.Z. and K.R. The three authors discussed the results and wrote the manuscript.

\section*{Competing interests}
The authors declare no competing interests.

\bibliography{references,new_refs}{}

%% =====================================================================
%% Supplementary Information -- appended at the end of the main text.
%% Numbering restarts with an "S" prefix.
%% =====================================================================
\clearpage
\FloatBarrier
\makeatletter
\setcounter{section}{0}\setcounter{subsection}{0}
\setcounter{figure}{0}\setcounter{table}{0}\setcounter{equation}{0}
\setcounter{secnumdepth}{1}
\renewcommand{\thesection}{\arabic{section}}
\renewcommand{\thefigure}{S\arabic{figure}}
\renewcommand{\thetable}{S\arabic{table}}
\renewcommand{\theequation}{S\arabic{equation}}
% unique hyperref anchors for the S-numbered objects
\renewcommand{\theHsection}{SN.\arabic{section}}
\renewcommand{\theHfigure}{S.\arabic{figure}}
\renewcommand{\theHtable}{S.\arabic{table}}
\renewcommand{\theHequation}{S.\arabic{equation}}
\def\SIfmt@section{Supplementary Note~\thesection:\enspace}
\def\SIfmt@subsection{\thesubsection\quad}
\renewcommand{\@seccntformat}[1]{\csname SIfmt@#1\endcsname}
\makeatother
\renewcommand{\figurename}{Supplementary Figure}
\renewcommand{\tablename}{Supplementary Table}

\begin{center}
{\LARGE\bfseries Supplementary Information}\\[8pt]
{\Large Exciton design via topological spin textures}\\[8pt]
{\normalsize Karim Rezouali, Selcuk S\"{o}zeri and Samir Lounis}
\end{center}
\vspace{1em}
\noindent This Supplementary Information provides the details behind the results presented in the main text. We first derive the Hamiltonian of a composite exciton in a spin texture (Supplementary Note~\ref{sn:model}) and discuss when the emergent flux is switched on and when the adiabatic description can be trusted (Supplementary Note~\ref{sn:switch}). We then present our first-principles results for the excitons of MoS$_2$ (Supplementary Note~\ref{sn:gw}), the structureless-spinor limit (Supplementary Note~\ref{sn:spinor}) and the weak-coupling regime, in which bright and dark excitons are mixed by the texture (Supplementary Note~\ref{sn:weak}). Supplementary Note~\ref{sn:pdfe} is devoted to Pd/Fe/Ir(111) as an illustration of a skyrmion host covered by an excitonic layer. The remaining notes collect additional results on the quantum-geometric trap (Supplementary Note~\ref{sn:trap}), on transport and single-exciton occupation (Supplementary Note~\ref{sn:transport}), and the material parameters used throughout (Supplementary Note~\ref{sn:materials}). Finally, Supplementary Note~\ref{sn:curves} lists, for every curve shown in the main text and in this Supplementary Information, the equation that was evaluated and the parameters that were used.

\tableofcontents
\clearpage

\section{Emergent fields of a composite exciton}
\label{sn:model}
Lengths are given in nm and energies in meV. We use the kinetic constant
\begin{equation}
t_M\equiv\frac{\hbar^2}{2M}=\frac{38.0998}{M/m_0}\ \text{meV\,nm}^2,
\label{Seq:tM}
\end{equation}
$\hbar=0.65821$\,meV\,ps and $k_BT=0.362$\,meV at 4.2\,K. Unless stated otherwise, the exciton mass is $M=1.01\,m_0$ (intralayer exciton of MoS$_2$, $m_e=0.47$, $m_h=0.54$) and $M=0.83\,m_0$ for interlayer excitons in MoS$_2$/WSe$_2$. The magnetization of a rotationally symmetric N\'eel skyrmion of charge $\nu$ is
\begin{equation}
\nn(\R)=\big(\sin\theta(R)\cos\nu\varphi,\ \sin\theta(R)\sin\nu\varphi,\ \cos\theta(R)\big),\qquad
\theta(R)=\pi+\sum_{\pm}\arcsin\tanh\frac{R\pm c}{w},
\label{Seq:profile}
\end{equation}
with $\theta(0)=\pi$ (core pointing down) and $\theta(\infty)=2\pi$ (up). The parameter $c$ is the skyrmion radius and $w$ the wall parameter. This 360$^\circ$-wall profile describes measured skyrmions accurately.

\paragraph{Model.} Near a valley $\tau=\pm1$, a conduction electron ($i=c$) or valence electron ($i=v$) of a transition-metal dichalcogenide placed next to the texture is described by
\begin{equation}
h_i=\frac{\mathbf p^2}{2m_i}+\mathbf b_i(\rr)\cdot\bm\sigma,\qquad
\mathbf b_i=J_i\,\nn(\rr)+\eta_i\tau\Delta_i\,\zz .
\label{Seq:hi}
\end{equation}
The first term of $\mathbf b_i$ is the exchange field of the texture, the second the Ising spin--orbit field, which points in opposite directions in the two valleys; $\eta_i=\pm1$ fixes the band ordering.

\paragraph{Direction of the effective field.} Because $\mathbf b_i$ is the sum of $J_i\nn$ and a field along $\zz$, it lies in the same vertical plane as $\nn$. Its azimuth is therefore that of the texture, $\phi_i=\nu\varphi$, and its polar angle $\Theta_i$ follows from
\begin{equation}
\cos\Theta_i=\frac{J_i\cos\theta+\eta_i\tau\Delta_i}{|\mathbf b_i|},\qquad
\sin\Theta_i=\frac{J_i\sin\theta}{|\mathbf b_i|},\qquad
|\mathbf b_i|=\sqrt{J_i^2+\Delta_i^2+2\eta_i\tau J_i\Delta_i\cos\theta}.
\label{Seq:Theta}
\end{equation}

\paragraph{Adiabatic projection.} We write the spinor of the branch parallel ($s_i=+1$) or antiparallel ($s_i=-1$) to $\mathbf b_i$. For $s_i=+1$,
\begin{equation}
\chi_i=\begin{pmatrix}\cos(\Theta_i/2)\\ e^{i\phi_i}\sin(\Theta_i/2)\end{pmatrix}.
\end{equation}
Projecting $h_i$ onto $\chi_i$, i.e., writing $\psi=\chi_i f$ and multiplying from the left by $\chi_i^\dagger$, gives
\begin{equation}
h_i^{\rm ad}=\frac{(\mathbf p-\hbar\mathbf a_i)^2}{2m_i}+\varepsilon_i+W_i,
\label{Seq:had}
\end{equation}
with the Berry connection, the branch energy and the geometric potential
\begin{align}
\mathbf a_i&=i\braket{\chi_i|\nabla\chi_i}=-\sin^2\frac{\Theta_i}{2}\,\nabla\phi_i=-\frac{1-\cos\Theta_i}{2}\,\nabla\phi_i,\label{Seq:ai}\\
\varepsilon_i&=s_i|\mathbf b_i|,\label{Seq:epsi}\\
W_i&=\frac{\hbar^2}{2m_i}\Big(\braket{\nabla\chi_i|\nabla\chi_i}-|\mathbf a_i|^2\Big)=\frac{\hbar^2}{8m_i}\Big[(\nabla\Theta_i)^2+\sin^2\Theta_i\,(\nabla\phi_i)^2\Big].\label{Seq:Wi}
\end{align}
The bracket in Eq.~\eqref{Seq:Wi} is the trace of the quantum metric of $\chi_i$, $g^{(i)}_{\mu\mu}=\mathrm{Re}\braket{\partial_\mu\chi_i|(1-\ket{\chi_i}\bra{\chi_i})|\partial_\mu\chi_i}$. Hence $W_i=(\hbar^2/2m_i)\,\mathrm{Tr}\,g^{(i)}$ is non-negative and largest where the effective field rotates quickly. The hole is the absence of a valence electron; its connection and energy are $\mathbf a_h=-\mathbf a_v$ and $\varepsilon_h=-\varepsilon_v$.

\paragraph{Emergent field and flux.} With $\nabla\phi_i=(\nu/R)\hat{\bm\varphi}$, Eq.~\eqref{Seq:ai} gives $a_{i,\varphi}=-\nu(1-\cos\Theta_i)/(2R)$, and the emergent field is
\begin{equation}
\Omega_i=\zz\cdot(\nabla\times\mathbf a_i)=\frac{1}{R}\frac{d}{dR}\big(R\,a_{i,\varphi}\big)=-\frac{\nu}{2R}\sin\Theta_i\,\frac{d\Theta_i}{dR}.
\end{equation}
Integrating over the plane, the flux in units of $\Phi_0=2\pi$ (in units where the emergent charge is one) is
\begin{equation}
\frac{\Phi_i}{\Phi_0}=\frac{1}{2\pi}\int d^2R\,\Omega_i=\frac{s_i\nu}{2}\Big[\cos\Theta_i(0)-\cos\Theta_i(\infty)\Big],
\label{Seq:flux}
\end{equation}
which is Eq.~\eqref{eq:fluxswitch} of the main text; the factor $s_i$ accounts for the branch.

\paragraph{Coordinates.} The two-body Hamiltonian in the adiabatic limit is
\begin{equation}
H=\sum_{i=e,h}\left[\frac{(\mathbf p_i-\hbar\mathbf a_i(\rr_i))^2}{2m_i}+\varepsilon_i(\rr_i)+W_i(\rr_i)\right]+V_C(\rr_e-\rr_h).
\end{equation}
We introduce $\R=\alpha\rr_e+\beta\rr_h$ and $\rr=\rr_e-\rr_h$, with $\alpha=m_e/M$, $\beta=m_h/M$, $M=m_e+m_h$ and $\mu=m_em_h/M$. Then $\rr_e=\R+\beta\rr$, $\rr_h=\R-\alpha\rr$, and the momenta are $\mathbf p_e=\alpha\mathbf P+\mathbf p$, $\mathbf p_h=\beta\mathbf P-\mathbf p$.

\paragraph{Expansion to first order in $\rr$.} We expand $\mathbf a_e(\R+\beta\rr)\simeq\mathbf a_e+\beta(\rr\cdot\nabla)\mathbf a_e$ and $\mathbf a_h(\R-\alpha\rr)\simeq\mathbf a_h-\alpha(\rr\cdot\nabla)\mathbf a_h$, all quantities on the right being evaluated at $\R$. The tensor $\partial_k a_j$ is split into symmetric and antisymmetric parts,
\begin{equation}
r_k\partial_ka_j=\tfrac12 r_k(\partial_ka_j+\partial_ja_k)+\tfrac12 r_k(\partial_ka_j-\partial_ja_k)
=\frac{\partial}{\partial r_j}\Big[\tfrac12 r_kr_l\partial_ka_l\Big]+\tfrac12\Omega\,(\zz\times\rr)_j .
\end{equation}
The symmetric part is a gradient with respect to $\rr$ and is removed by a gauge transformation in the relative coordinate; it contributes only at second order in $\rr$ to the centre-of-mass motion, which we neglect. The kinetic momenta become
\begin{align}
\bm\pi_e&=\alpha\mathbf P+\mathbf p-\hbar\mathbf a_e-\tfrac{\hbar\beta}{2}\Omega_e\,\zz\times\rr,\\
\bm\pi_h&=\beta\mathbf P-\mathbf p-\hbar\mathbf a_h+\tfrac{\hbar\alpha}{2}\Omega_h\,\zz\times\rr .
\end{align}

\paragraph{Exact separation of the kinetic energy.} For any pair of momenta, $\bm\pi_e^2/2m_e+\bm\pi_h^2/2m_h=\bm\Pi^2/2M+\bm\pi^2/2\mu$ with $\bm\Pi=\bm\pi_e+\bm\pi_h$ and $\bm\pi=\beta\bm\pi_e-\alpha\bm\pi_h$. Here
\begin{align}
\bm\Pi&=\mathbf P-\hbar\mathbf a_X-\tfrac{\hbar}{2}\Omega_-\,\zz\times\rr,\qquad \mathbf a_X=\mathbf a_e+\mathbf a_h,\quad \Omega_-=\beta\Omega_e-\alpha\Omega_h,\\
\bm\pi&=\mathbf p-\hbar\mathbf c-\tfrac{\hbar}{2}\Omega_+\,\zz\times\rr,\qquad \mathbf c=\beta\mathbf a_e-\alpha\mathbf a_h,\quad \Omega_+=\beta^2\Omega_e+\alpha^2\Omega_h .
\end{align}

\paragraph{Gauge transformation.} The $\rr$-independent potential $\mathbf c(\R)$ in the relative momentum is removed by $\psi=U\tilde\psi$ with $U=\exp[i\,\rr\cdot\mathbf c(\R)]$. Since $U$ depends on $\R$, it also shifts the centre-of-mass momentum, $U^\dagger\mathbf PU=\mathbf P+\hbar\nabla_{\R}(\rr\cdot\mathbf c)$. Using $\nabla(\rr\cdot\mathbf c)=(\rr\cdot\nabla)\mathbf c+\rr\times(\nabla\times\mathbf c)$ at fixed $\rr$, and $\nabla\times\mathbf c=\Omega_-\zz$, we find $\nabla_{\R}(\rr\cdot\mathbf c)=[\text{gradient in }\rr]-\tfrac12\Omega_-\,\zz\times\rr$. The antisymmetric part therefore doubles the coefficient of $\zz\times\rr$ in $\bm\Pi$, and we obtain Eq.~\eqref{eq:H2body} of the main text,
\begin{equation}
H=\frac{\big[\mathbf P-\hbar\mathbf a_X-\hbar\Omega_-\,\zz\times\rr\big]^2}{2M}+\frac{\big[\mathbf p-\tfrac{\hbar}{2}\Omega_+\,\zz\times\rr\big]^2}{2\mu}+V_C(\rr)+\sum_{i=e,h}(\varepsilon_i+W_i).
\label{Seq:H2}
\end{equation}

\paragraph{Gor'kov--Dzyaloshinskii limit.} For a uniform magnetic field, $\hbar\mathbf a_{e}=-e\mathbf A$ and $\hbar\mathbf a_h=+e\mathbf A$, so that $\mathbf a_X=0$, $\hbar\Omega_-=-eB$ and $\hbar\Omega_+=-eB(m_h-m_e)/M$. Equation~\eqref{Seq:H2} then reduces to the Hamiltonian of an exciton in a magnetic field. In a skyrmion, $\mathbf a_e$ and $\mathbf a_h$ do not cancel, because each depends on its own $J_i/\Delta_i$, spin and valley.

\paragraph{Averaging over the 1s state.} For an exciton small compared with the wall, we average Eq.~\eqref{Seq:H2} over the 1s state $\phi_{1s}(\rr)$. Terms linear in $\rr$ vanish by symmetry, and so does the orbital Zeeman term $\propto\Omega_+(\zz\times\rr)\cdot\mathbf p$ for an $s$ state. With $\langle|\zz\times\rr|^2\rangle=\langle r^2\rangle$,
\begin{equation}
H_{\rm CM}=\frac{(\mathbf P-\hbar\mathbf a_X)^2}{2M}+\sum_i(\varepsilon_i+W_i)+\hbar^2\langle r^2\rangle\left[\frac{\Omega_-^2}{2M}+\frac{\Omega_+^2}{8\mu}\right].
\label{Seq:HCM}
\end{equation}
The last term is the emergent diamagnetic shift. The first, second (geometric) and last terms are non-negative; only the exchange energies $\varepsilon_i$ can produce an attractive well.

\section{The flux switch and the validity of the adiabatic picture}
\label{sn:switch}
From Eq.~\eqref{Seq:Theta}, the $z$ component of $\mathbf b_i$ is $J_i+\eta_i\tau\Delta_i$ far from the skyrmion ($\theta=2\pi$) and $-J_i+\eta_i\tau\Delta_i$ in the core ($\theta=\pi$), while $\sin\Theta_i=0$ in both places. Therefore $\cos\Theta_i(0)=\mathrm{sign}(-J_i+\eta_i\tau\Delta_i)$ and $\cos\Theta_i(\infty)=\mathrm{sign}(J_i+\eta_i\tau\Delta_i)$. The two signs differ only if $J_i>\Delta_i$, and Eq.~\eqref{Seq:flux} gives
\begin{equation}
\frac{\Phi_i}{\Phi_0}=\begin{cases}\pm\nu, & J_i>\Delta_i,\\ 0, & J_i<\Delta_i.\end{cases}
\label{Seq:switch}
\end{equation}
For $J_i<\Delta_i$, $\mathbf b_i$ tilts in the wall and returns to its original direction, so the swept solid angle vanishes. For $J_i>\Delta_i$, it is reversed in the core and sweeps the full sphere (main text Fig.~\ref{fig:bands}b). Because of spin--valley locking, the band-edge carrier occupies opposite branches in $K$ and $K'$, so that $s_i$, and with it the flux, has opposite signs in the two valleys.

\paragraph{Validity.} The projection requires the local splitting $2|\mathbf b_i|$ to exceed the kinetic energy associated with the rotation of $\hat{\mathbf b}_i$, $\hbar^2/(2m_iw^2)$. Near $J_i\approx\Delta_i$, $|\mathbf b_i|$ becomes small in the core and the condition fails. The adiabatic description is quantitative when $J_i>\max\{\Delta_i,\hbar^2/(2m_iw^2)\}$. For a mass $m_0$, $\hbar^2/(2m_0w^2)=38.1/w^2$\,meV, i.e., 16.9\,meV for $w=1.5$\,nm and 1.52\,meV for $w=5$\,nm.

\section{Excitons of monolayer MoS$_2$ from first principles}
\label{sn:gw}
To anchor the model parameters, we computed the excitons of free-standing monolayer MoS$_2$ within many-body perturbation theory. Ground-state calculations were performed with \textsc{Quantum Espresso}~\cite{qe,giannozzi2009} using ONCV pseudopotentials~\cite{hamann2013}, the PBE functional~\cite{perdew1996} and spin--orbit coupling. Quasiparticle energies were obtained within $G_0W_0$~\cite{Hybertsen1985}, with the generalized plasmon-pole model~\cite{godby1989} and a truncated Coulomb interaction~\cite{rozzi2006}, and the Bethe--Salpeter equation was solved with \textsc{Yambo}~\cite{marini2009} on a $27\times27\times1$ $k$-mesh. Since spin--orbit coupling is included throughout, bright and spin-forbidden dark excitons are obtained on the same footing.

The results are summarized in Supplementary Fig.~\ref{fig:S1}. The quasiparticle gap at $K$ is 2.73\,eV, and the spin--orbit splittings at $K$ are 149.5\,meV in the valence band and 1.6\,meV in the conduction band. The bright A and B excitons appear at 1.91 and 2.03\,eV, so that the A exciton is bound by 0.82\,eV in the free-standing layer. The most relevant result for the present work concerns the fine structure. The lowest exciton is not the bright A exciton but a spin-forbidden dark state 23\,meV below it, with an oscillator strength smaller by about six orders of magnitude. A similar dark state lies 15\,meV below the bright B exciton. The ordering agrees with magneto-optical experiments on hBN-encapsulated MoS$_2$, which find the dark exciton 14\,meV below the bright one~\cite{robert2020}. The smaller experimental value is consistent with the additional screening by hBN, which reduces the short-range electron--hole exchange that contributes to the bright--dark splitting. The real-space density of the A exciton (Supplementary Fig.~\ref{fig:S1}c) extends over about a nanometre, in line with the radius of 1.2\,nm inferred from high-field magneto-optics~\cite{goryca2019}.

\begin{figure}[p]\centering\includegraphics[width=\textwidth]{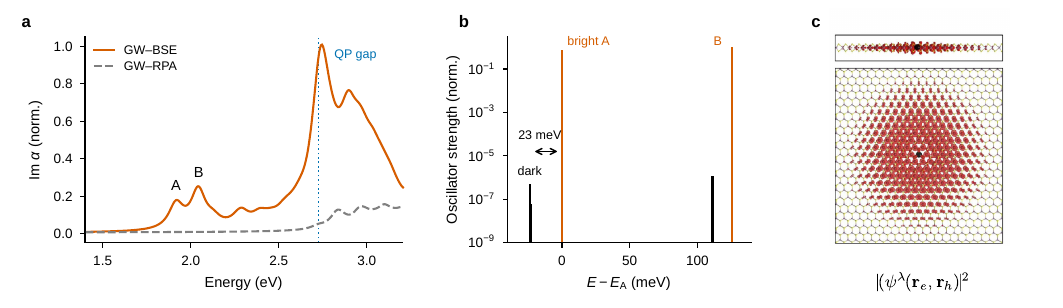}
\caption{\textbf{Excitons of free-standing monolayer MoS$_2$ from $GW$--BSE.} \textbf{a}, Imaginary part of the in-plane polarizability with (solid) and without (dashed) electron--hole interaction, both on top of $GW$ quasiparticle energies; A and B mark the bright 1s excitons at 1.91 and 2.03\,eV, the dotted line the quasiparticle gap at $K$ (2.73\,eV). \textbf{b}, Oscillator strengths of the lowest exciton states relative to the bright A exciton: the spin-forbidden dark A exciton lies 23\,meV below the bright one, and the dark B exciton 15\,meV below the bright B exciton. \textbf{c}, Real-space electron density of the A exciton with the hole fixed near a Mo atom (top: side view; bottom: top view).}\label{fig:S1}\end{figure}

\section{The structureless-spinor limit}
\label{sn:spinor}
When every constituent winds with the texture, the exciton can be described as a structureless spinor with a single exchange coupling, as in ref.~\cite{kazemi2023}. In this limit, the emergent vector potential has a long-range tail, $\mathbf a\propto\nu/\rho$, which makes the scattering problem delicate: the result depends on whether the tail is treated exactly or perturbatively, and on the gauge in which the calculation is performed. To clarify this point, we solved the structureless model exactly in partial waves, including the geometric scalar potential, in the gauge regular at the origin and in the gauge regular at infinity. Both gauges give identical cross-sections, and the results are converged in the number of partial waves and in the matching radius.

\begin{figure}[p]\centering\includegraphics[width=0.66\textwidth]{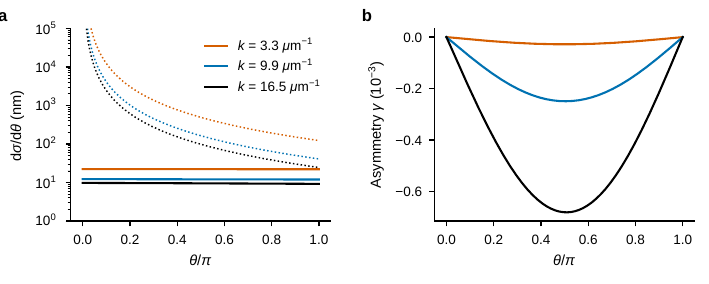}
\caption{\textbf{Gauge-consistent exact solution of the structureless-spinor model.} Model of ref.~\cite{kazemi2023} including the geometric scalar potential, with exciton mass $5.6\,m_0$, skyrmion radius $\approx5$\,nm and the wave vectors of that work. \textbf{a}, Exact differential cross-section (solid) compared with the small-angle form $4/(k\theta^2)$ obtained when the $1/\rho$ tail of the vector potential is treated perturbatively (dotted). \textbf{b}, Exact left--right asymmetry. Results are identical in the gauges regular at the origin and at infinity, and are converged in partial waves and matching radius.}\label{fig:S2}\end{figure}

For the parameters of ref.~\cite{kazemi2023} (exciton mass $5.6\,m_0$, skyrmion radius of about 5\,nm, wave vectors between 3.3 and 16.5\,$\mu$m$^{-1}$), the exact differential cross-section is smooth and nearly isotropic (Supplementary Fig.~\ref{fig:S2}a). It does not show the forward divergence $4/(k\theta^2)$ that results when the $1/\rho$ tail is treated to first order. The left--right asymmetry remains below $10^{-3}$ for all angles (Supplementary Fig.~\ref{fig:S2}b). This is what one expects physically. At these wave vectors $kR\ll1$, the exciton cannot resolve the texture, and chirality in the scattering requires $kR\gtrsim1$, as in the calculations of Fig.~\ref{fig:switch} of the main text.

\section{Bright and dark excitons: from weak to strong coupling}
\label{sn:weak}
\paragraph{Local Hamiltonian.} When the hole spin is pinned by the large valence splitting, the texture acts on the exciton through the electron spin, which distinguishes bright ($B$, electron spin parallel to the hole) from dark ($D$) states. Measuring energies from the bright exciton in zero field, the local Hamiltonian in valley $\tau$ is
\begin{equation}
H_\tau(\R)=\begin{pmatrix}\tau J_c\,n_z^{e}-\tau J_v\,n_z^{h} & J_c\,(n_x^{e}-i\tau n_y^{e})\\ J_c\,(n_x^{e}+i\tau n_y^{e}) & -\tau J_c\,n_z^{e}-\tau J_v\,n_z^{h}-\Delta_\mathrm{BD}\end{pmatrix}.
\label{Seq:Htau}
\end{equation}
Here $\mathbf n^{e}$ and $n_z^{h}$ are the texture seen by the electron and the hole. For an exciton small compared with the skyrmion, both are the local texture at $\R$, $n_z=\cos\theta$ and $n_x-i\tau n_y=\sin\theta\,e^{-i\tau\nu\varphi}$. This gives
\begin{equation}
H_\tau(\R)=\begin{pmatrix}\tau(J_c-J_v)\cos\theta & J_c\sin\theta\,e^{-i\tau\nu\varphi}\\ J_c\sin\theta\,e^{i\tau\nu\varphi} & -\Delta_\mathrm{BD}-\tau(J_c+J_v)\cos\theta\end{pmatrix}.
\label{Seq:Htau2}
\end{equation}
In a uniform magnetization ($\cos\theta=1$), the bright exciton is split between the valleys by $2(J_c-J_v)$.

\paragraph{Pseudospin form and threshold.} Writing $H_\tau=d_0+\mathbf b\cdot\bm\sigma$ in the bright--dark pseudospin space gives
\begin{equation}
d_0=-\tau J_v\cos\theta-\tfrac{\Delta_\mathrm{BD}}{2},\qquad b_z=\tau J_c\cos\theta+\tfrac{\Delta_\mathrm{BD}}{2},\qquad b_x-ib_y=J_c\sin\theta\,e^{-i\tau\nu\varphi}.
\end{equation}
This is Eq.~\eqref{Seq:hi} for a pseudospin, with exchange $J_c$ and Ising term $\Delta_\mathrm{BD}/2$. Outside the skyrmion $b_z=\tau J_c+\Delta_\mathrm{BD}/2$, in the core $b_z=-\tau J_c+\Delta_\mathrm{BD}/2$. The pseudospin field is reversed, and the branch connected to the bright exciton carries a full flux quantum, if and only if
\begin{equation}
J_c>J^*=\frac{|\Delta_\mathrm{BD}|}{2},
\label{Seq:Jstar}
\end{equation}
which is the threshold used in Fig.~\ref{fig:map}c for intralayer excitons. For interlayer excitons the electron--hole exchange is weak and the threshold reduces to the single-particle condition $J_c>\Delta_c$ of Supplementary Note~\ref{sn:switch}.

\paragraph{Angular-momentum selection rule.} Writing $\psi_B=R^{-1/2}u_B(R)e^{i\ell\varphi}$, the off-diagonal term $e^{i\tau\nu\varphi}$ of Eq.~\eqref{Seq:Htau2} couples it only to $\psi_D=R^{-1/2}u_D(R)e^{i(\ell+\tau\nu)\varphi}$. Each scattering event therefore transfers $\tau\nu$ units of angular momentum.

\paragraph{Weak-coupling ring.} For $J_c\ll|\Delta_\mathrm{BD}|$, first-order perturbation theory gives the bright admixture of the dark branch, $|v_B|^2\simeq(J_c\sin\theta/\Delta_\mathrm{BD})^2$, which is maximal in the wall. The exact maximum over the wall ($\theta=\pi/2$, diagonal terms set to the bare splitting) is
\begin{equation}
|v_B|^2_{\max}=\frac12\left[1-\frac{\Delta_\mathrm{BD}}{\sqrt{\Delta_\mathrm{BD}^2+4J_c^2}}\right].
\label{Seq:ring}
\end{equation}

\begin{figure}[p]\centering\includegraphics[width=0.72\textwidth]{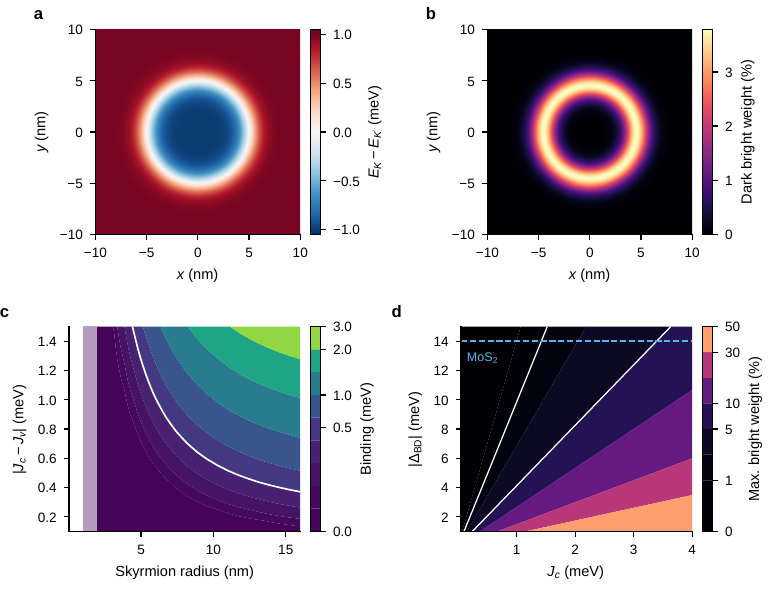}
\caption{\textbf{Weak-coupling fingerprints of a single skyrmion} (radius 5\,nm, wall 1\,nm, $a_\mathrm{B}=1.5$\,nm, $J_c=1$\,meV, $J_v=0.5$\,meV, $\Delta_\mathrm{BD}=5$\,meV). \textbf{a}, Local valley splitting of the bright exciton. \textbf{b}, Bright weight acquired by the dark exciton. \textbf{c}, Binding energy in the valley-odd well versus skyrmion radius and $|J_c-J_v|$; the white line marks $k_\mathrm{B}T$ at 4.2\,K and the shaded strip Pd/Fe/Ir(111). \textbf{d}, Maximum bright weight of the dark exciton versus $J_c$ and $|\Delta_\mathrm{BD}|$, with 1\% and 5\% contours; the dashed line marks MoS$_2$~\cite{robert2020}.}\label{fig:S4}\end{figure}

Three consequences follow directly from Eq.~\eqref{Seq:Htau2}. First, the diagonal terms form a scalar landscape that is odd in the valley index: a skyrmion creates a region in its core where the valley ordering of the bright exciton is reversed (Supplementary Fig.~\ref{fig:S4}a). Second, the off-diagonal term mixes the dark exciton into the bright one, and vice versa, where the magnetization lies in plane. For $J_c\ll|\Delta_{BD}|$, the dark exciton acquires a bright weight of about $(J_c\sin\theta/\Delta_{BD})^2$, which is largest in the wall and gives a ring of brightened dark excitons (Supplementary Fig.~\ref{fig:S4}b,d). Third, the phase $e^{-i\tau\nu\varphi}$ imposes the selection rule $\ell_D=\ell_B+\tau\nu$ on the angular momenta of the coupled channels.

\paragraph{Averaging over the internal motion of the exciton.} For skyrmions comparable in size to the exciton, the electron at $\R+\beta\rr$ and the hole at $\R-\alpha\rr$ see different parts of the texture. The fields in Eq.~\eqref{Seq:Htau} are then
\begin{equation}
\bar{\mathbf n}^{e}(\R)=\int d^2r\,|\phi_{1s}(\rr)|^2\,\nn(\R+\beta\rr),\qquad
\bar n^{h}_z(\R)=\int d^2r\,|\phi_{1s}(\rr)|^2\,n_z(\R-\alpha\rr),
\end{equation}
with the two-dimensional 1s density $|\phi_{1s}|^2\propto e^{-4r/a_B}$. Equivalently, $\bar{\mathbf n}^{e}$ is the texture convolved with $e^{-4|\mathbf s|/(\beta a_B)}$ and $\bar n^h_z$ with $e^{-4|\mathbf s|/(\alpha a_B)}$ (normalized kernels). The local eigenvalues and eigenvectors of Eq.~\eqref{Seq:Htau} with these fields give the maps of Supplementary Figs.~\ref{fig:S4} and \ref{fig:S6}.

\paragraph{The adiabatic exciton.} Diagonalizing Eq.~\eqref{Seq:Htau2} at each $R$ (at $\varphi=0$) and following the eigenvector $v(R)=(v_B,v_D)$ that is bright at infinity, the projection of Supplementary Note~\ref{sn:model} applied to the pseudospin gives the single-channel radial problem
\begin{equation}
-u''+\left[\frac{(\ell-A(R))^2-\frac14}{R^2}+\frac{E_{\rm br}(R)-E_{\rm br}(\infty)+W(R)-\varepsilon}{t_M}\right]u=0,
\label{Seq:adradial}
\end{equation}
with the Berry potential (flux quanta enclosed within $R$), in the gauge regular at the origin,
\begin{equation}
A_{\rm reg}(R)=-\tau\nu\big[|v_D(R)|^2-|v_D(0)|^2\big],
\end{equation}
and in the gauge regular at infinity, used in the calculations, $A(R)=A_{\rm reg}(R)-A_{\rm reg}(\infty)$. The geometric potential is
\begin{equation}
W(R)=t_M\left[\Big|\frac{dv}{dR}\Big|^2+\nu^2\frac{|v_B|^2|v_D|^2}{R^2}\right].
\end{equation}
The net flux of the branch is $\Phi/\Phi_0=A_{\rm reg}(\infty)=\tau\nu|v_D(0)|^2$, i.e., $0$ below the threshold of Eq.~\eqref{Seq:Jstar} and $\tau\nu$ above it.

For a texture that the exciton follows completely ($|v_D|^2=\sin^2(\theta/2)$ along the profile), these expressions reduce to
\begin{equation}
A(R)=\frac{\nu}{2}\big[\cos\theta(R)-\cos\theta(\infty)\big],\qquad
W(R)=\frac{t_M}{4}\left[\theta'^2+\nu^2\frac{\sin^2\theta}{R^2}\right],
\label{Seq:AWfull}
\end{equation}
and the effective radial potential becomes
\begin{equation}
U_\ell(R)=t_M\frac{(\ell-A(R))^2-\frac14}{R^2}+W(R)+V(R),\qquad V(R)=-V_0\,\frac{1-\cos\theta(R)}{2},
\label{Seq:Ul}
\end{equation}
where $V$ is an optional exchange well of depth $V_0$ in the core. In the core $A=-\nu$, so the channel $\ell=-\nu$ is $s$-like inside and meets the centrifugal barrier and $W$ outside.

\paragraph{Exact two-channel scattering.}
\paragraph{Radial equations.} With $\psi_B$ and $\psi_D$ as above and $m_B=\ell$, $m_D=\ell+\tau\nu$, the Schr\"odinger equation for $-t_M\nabla^2+H_\tau$ at energy $E$ becomes
\begin{align}
u_B''&=\left[\frac{m_B^2-\frac14}{R^2}+\frac{E_B(R)-E}{t_M}\right]u_B+\frac{J_c\sin\theta}{t_M}\,u_D,\\
u_D''&=\left[\frac{m_D^2-\frac14}{R^2}+\frac{E_D(R)-E}{t_M}\right]u_D+\frac{J_c\sin\theta}{t_M}\,u_B,
\end{align}
with $E_B=\tau(J_c-J_v)\cos\theta$ and $E_D=-\Delta_\mathrm{BD}-\tau(J_c+J_v)\cos\theta$. The channel momenta are $k_c=[(E-E_c(\infty))/t_M]^{1/2}$, with $E_B(\infty)=\tau(J_c-J_v)$ and $E_D(\infty)=-\Delta_\mathrm{BD}-\tau(J_c+J_v)$. The incident bright exciton has kinetic energy $\varepsilon$, i.e., $E=\tau(J_c-J_v)+\varepsilon$, and we use $\varepsilon=k_BT=0.36$\,meV unless stated otherwise.

\paragraph{S matrix.} Two regular solutions, $u_c\propto R^{|m_c|+1/2}$ near the origin in one channel and zero in the other, are integrated outward. At large $R$, each component is written as $u_c=A_c\sqrt R\,H^{(2)}_{m_c}(k_cR)+B_c\sqrt R\,H^{(1)}_{m_c}(k_cR)$. Since $\sqrt R\,H^{(1,2)}_m(kR)$ has an amplitude proportional to $k^{-1/2}$, the outgoing flux in channel $c$ is proportional to $|B_c|^2$ independently of $k_c$. The matrix $S=BA^{-1}$ is therefore flux-normalized and unitary for open channels; we checked $\max|S^\dagger S-1|<10^{-3}$.

\paragraph{Cross-sections.} The elastic and conversion amplitudes are
\begin{equation}
f_{BB}(\theta_s)=\frac{1}{\sqrt{2\pi k_B}}\sum_\ell\big(S^{\ell}_{BB}-1\big)e^{i\ell\theta_s},\qquad
f_{DB}(\theta_s)=\frac{1}{\sqrt{2\pi k_B}}\sum_\ell S^{\ell}_{DB}\,e^{i(\ell+\tau\nu)\theta_s},
\end{equation}
and
\begin{equation}
\sigma_{\rm el}=\int d\theta_s\,|f_{BB}|^2=\frac{1}{k_B}\sum_\ell|S^\ell_{BB}-1|^2,\qquad
\sigma_{B\to D}=\frac{1}{k_B}\sum_\ell|S^\ell_{DB}|^2 .
\end{equation}
The skew and the left--right asymmetry are
\begin{equation}
\langle\sin\theta_s\rangle=\frac{\int d\theta_s\,\sin\theta_s\,|f_{BB}|^2}{\int d\theta_s\,|f_{BB}|^2},\qquad
\gamma(\theta_s)=\frac{|f_{BB}(\theta_s)|^2-|f_{BB}(-\theta_s)|^2}{|f_{BB}(\theta_s)|^2+|f_{BB}(-\theta_s)|^2}.
\end{equation}
In the adiabatic (gauge-field) description, the same formulas are used with a single channel, $S^\ell=b_\ell/a_\ell$ from Eq.~\eqref{Seq:adradial}. For $K'$ above the flux switch,
the dark exciton lies above the bright one and the dark channel is closed at the energies considered ($k_D^2<0$, i.e., $J_c/\Delta_\mathrm{BD}>0.54$ for $\varepsilon=k_BT$). In this case, the two regular solutions are integrated outward to $R=c$, while the decaying dark solution $\sqrt R\,K_{|m_D|}(\kappa_DR)$ and the two real bright solutions $\sqrt R\,J_{|m_B|}(k_BR)$ and $\sqrt R\,Y_{|m_B|}(k_BR)$ are integrated inward from $R_{\max}$, the stable direction for the closed channel, with Gram--Schmidt re-orthogonalization. Matching at $R=c$ gives $S_{BB}$ with $|S_{BB}|=1$ and $\sigma_{B\to D}=0$.
%the dark channel becomes closed and would require closed-channel matching, which is not implemented; only $K$ is shown for the exact solution in Fig.~3.

\begin{figure}[p]\centering\includegraphics[width=\textwidth]{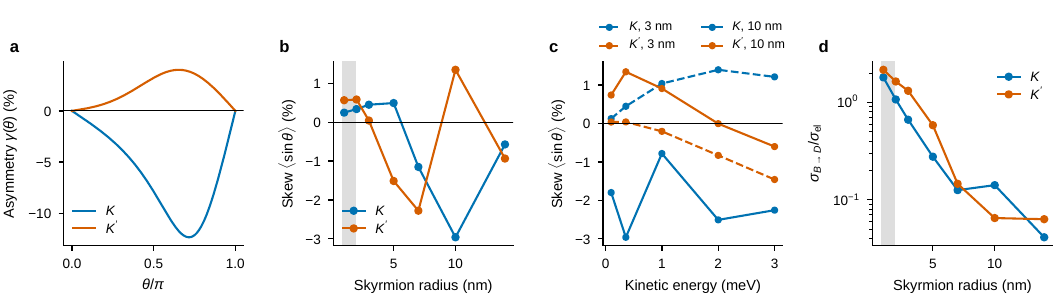}
\caption{\textbf{Chiral scattering in the weak-coupling limit} (wall width 1\,nm, $J_c=1$\,meV, $J_v=0.5$\,meV, $\Delta_\mathrm{BD}=5$\,meV, kinetic energy $k_\mathrm{B}T$ at 4.2\,K unless stated). \textbf{a}, Left--right asymmetry of elastic scattering by a 10\,nm skyrmion. \textbf{b}, Skew versus skyrmion radius; the sign reverses for antiskyrmions. \textbf{c}, Skew versus kinetic energy for radii of 3 and 10\,nm. \textbf{d}, Ratio of bright-to-dark conversion to elastic scattering. Grey bands mark Pd/Fe/Ir(111).}\label{fig:S3}\end{figure}

In the weak-coupling regime, the scattering is weak but chiral (Supplementary Fig.~\ref{fig:S3}). The asymmetry reverses for an antiskyrmion and has opposite sign in the two valleys, it changes sign as a function of the skyrmion radius and of the kinetic energy, and it is accompanied by a substantial conversion of bright into dark excitons, which becomes comparable to elastic scattering for the smallest skyrmions.

\paragraph{Bound states in the valley-odd landscape.} The diagonal of Eq.~\eqref{Seq:Htau2} also produces an attractive well in one valley when $J_c\neq J_v$. Its depth is $2|J_c-J_v|$ for a fully reversed core. We computed the binding energy of the lowest state as a function of the skyrmion radius and of $|J_c-J_v|$ on a cell-centred finite-volume grid (Supplementary Fig.~\ref{fig:S4}c). Binding above $k_BT$ at 4.2\,K requires $|J_c-J_v|$ of about 1\,meV for a skyrmion radius of 5\,nm, and about 0.4\,meV for a radius of 15\,nm. Larger skyrmions are therefore clearly favourable.

\section{Pd/Fe/Ir(111) as an illustrative skyrmion host}
\label{sn:pdfe}
\begin{figure}[p]\centering\includegraphics[width=\textwidth]{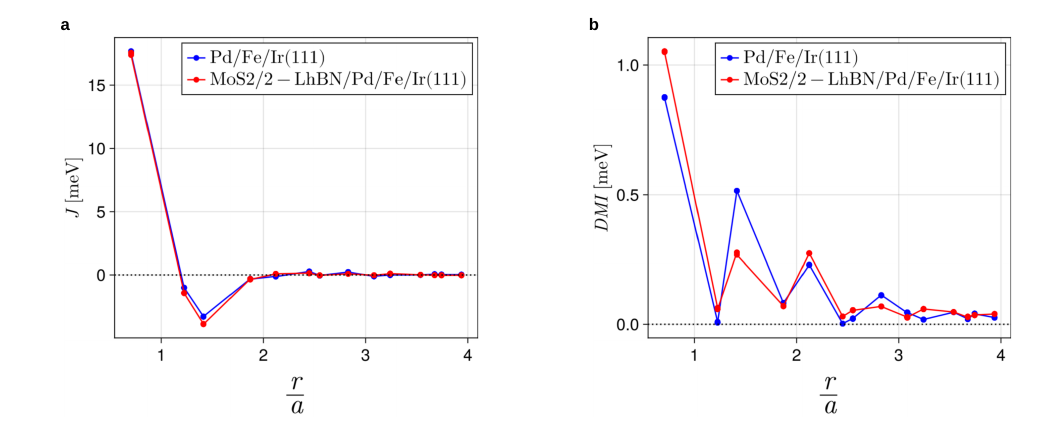}
\caption{\textbf{Skyrmion-stabilizing interactions survive a van der Waals overlayer.} \textbf{a}, Isotropic exchange $J_{ij}$ and \textbf{b}, magnitude of the Dzyaloshinskii--Moriya interaction between Fe atoms versus distance, for bare Pd/Fe/Ir(111) (blue) and for Pd/Fe/Ir(111) covered by two hBN layers and a MoS$_2$ monolayer (red), from first-principles KKR calculations.}\label{fig:S5}\end{figure}

Pd/Fe/Ir(111) is the system in which individual skyrmions were first written and deleted with a spin-polarized tip~\cite{romming2013}, and its micromagnetic parameters are well established~\cite{romming2015}. We use it here for two purposes. The first is to test whether a skyrmion host keeps its skyrmions when it is covered by a van der Waals overlayer. To this end, we computed from first principles the magnetic interactions of Pd/Fe/Ir(111) without and with an overlayer of two hBN layers and a MoS$_2$ monolayer. The isotropic exchange interactions $J_{ij}$ and the Dzyaloshinskii--Moriya interactions between the Fe atoms are essentially unchanged by the overlayer (Supplementary Fig.~\ref{fig:S5}). Since these interactions determine the size and the stability of the skyrmions, we conclude that the skyrmions survive the van der Waals overlayer. We stress that we use these calculations only for this purpose. The relaxed geometry does not include van der Waals corrections, and it would not be reliable enough for quantitative values of the proximity exchange acting on MoS$_2$.

The second purpose is to illustrate the small-size limit of our theory. Minimizing the micromagnetic functional with the parameters of ref.~\cite{romming2015} ($A=2.0$\,pJ/m, $D=3.9$\,mJ/m$^2$, $K=2.5$\,MJ/m$^3$, $M_s=1.1$\,MA/m) reproduces the measured skyrmion sizes, with radii between about 1.9 and 0.8\,nm for fields between 1 and 3\,T (Supplementary Fig.~\ref{fig:S6}a). These skyrmions are comparable in size to the exciton, so that the excitonic fingerprints of Supplementary Note~\ref{sn:weak} are averaged over the extent of the exciton. We find that the valley reversal in the core and the ring of brightened dark excitons keep between about 55 and 98\% of their large-skyrmion values, depending on field and exciton radius (Supplementary Fig.~\ref{fig:S6}b,c). Bright-to-dark conversion is comparable to elastic scattering in this regime (Supplementary Fig.~\ref{fig:S6}d). Pd/Fe/Ir(111) is metallic, however, and a spacer is needed to protect the exciton. For the excitonic coupling itself, insulating skyrmion hosts in direct contact with the excitonic layer are therefore preferable, as discussed in the main text.

\paragraph{Micromagnetic profiles.} The energy of a rotationally symmetric skyrmion in a film with exchange stiffness $A$, interfacial DMI $D$, uniaxial anisotropy $K$, saturation magnetization $M_s$ and field $B$ is, per unit thickness,
\begin{equation}
E[\theta]=2\pi\int_0^\infty\rho\,d\rho\left[A\Big(\theta'^2+\frac{\sin^2\theta}{\rho^2}\Big)-D\Big(\theta'+\frac{\sin\theta\cos\theta}{\rho}\Big)+K\sin^2\theta-M_sB(\cos\theta-1)\right],
\end{equation}
measured from the ferromagnetic state. We minimize it within the profile $\theta(\rho)=\pi+\sum_\pm\arcsin\tanh[(\rho\pm c)/(w_R/2)]$ with respect to $c$ and the wall width $w_R$, using $A=2.0$\,pJ/m, $D=3.9$\,mJ/m$^2$, $K=2.5$\,MJ/m$^3$ and $M_s=1.1$\,MA/m. The skyrmion diameter is twice the radius at which $n_z=0$. In the model of Eq.~\eqref{Seq:profile}, this profile corresponds to $w=w_R/2$.

\begin{figure}[p]\centering\includegraphics[width=0.72\textwidth]{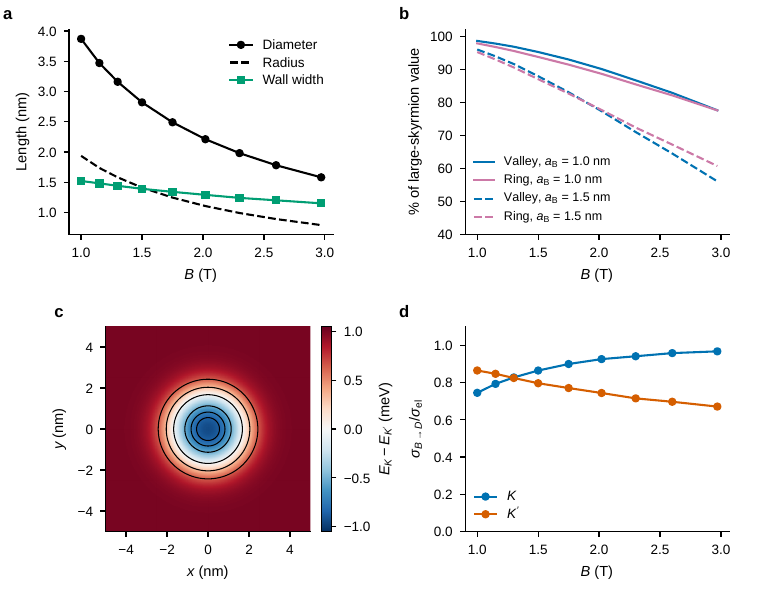}
\caption{\textbf{Excitonic fingerprints of nanometric skyrmions: the example of Pd/Fe/Ir(111).} \textbf{a}, Skyrmion diameter, radius and wall width versus field from the micromagnetic functional with the parameters of ref.~\cite{romming2015}, reproducing the measured sizes. \textbf{b}, Core valley reversal and ring weight relative to their large-skyrmion values, for exciton radii of 1.0 and 1.5\,nm. \textbf{c}, Valley-splitting map at 1.15\,T with contours of dark-exciton bright weight (1, 2 and 3\%). \textbf{d}, Ratio of bright-to-dark conversion to elastic scattering versus field. Parameters as in Supplementary Fig.~\ref{fig:S4}.}\label{fig:S6}\end{figure}

\section{The quantum-geometric trap}
\label{sn:trap}
In the main text we showed, using control calculations, that the barrier of the trap originates from the quantum metric, while the topology shifts the angular momentum of the trapped state. Here we summarize how the trap depends on the topological charge and on the size of the skyrmion, and what happens when an attractive exchange well is added.

\paragraph{Phase shifts.} For $V_0=0$, Eq.~\eqref{Seq:Ul} is integrated outward from $u\propto R^{|\ell-A(0)|+1/2}$ and matched at $R_{\max}$ to $\sqrt R\,[a\,J_{|\ell|}(kR)+b\,Y_{|\ell|}(kR)]$, $k=(E/t_M)^{1/2}$. Because $A(R\to\infty)=0$ in this gauge, the asymptotic channel is free with angular momentum $\ell$. The phase shift is $\delta_\ell=\arg(a-ib)$, i.e., $\tan\delta_\ell=-b/a$.

\paragraph{Time delay and lifetime.} Resonances are identified from the Wigner time delay
\begin{equation}
\tau_W(E)=2\hbar\,\frac{d\delta_\ell}{dE},
\label{Seq:tauW}
\end{equation}
with width $\Gamma=4\hbar/\tau_W^{\max}$, lifetime $\hbar/\Gamma=\tau_W^{\max}/4$ and quality factor $Q=E_r/\Gamma$.

\paragraph{Control calculations.} (i) ``No flux'': $A\equiv0$, keeping $W$. (ii) Skyrmionium: $\theta(R)=\tfrac{\pi}{2}\big[2+\tanh\tfrac{R-c_1}{w}+\tanh\tfrac{R-c_2}{w}\big]$, which winds from $0$ through $\pi$ back to $2\pi$ and has zero net flux. (iii) Linear wall: $\theta(R)=\pi+\pi\,\mathrm{clip}\big(\tfrac{R-(c-w)}{2w},0,1\big)$.

\paragraph{Bound states.} With $V_0>0$, the radial equation for $\psi(R)$ is discretized on cell centres $R_j=(j-\tfrac12)h$, $j=1,\dots,N$, with $R_{j\pm1/2}=R_j\pm h/2$. The finite-volume Laplacian gives a tridiagonal symmetric matrix with
\begin{equation}
H_{jj}=t_M\frac{R_{j+1/2}+R_{j-1/2}}{h^2R_j}+t_M\frac{(\ell-A(R_j))^2}{R_j^2}+W(R_j)+V(R_j),\qquad
H_{j,j+1}=-t_M\frac{R_{j+1/2}}{h^2\sqrt{R_jR_{j+1}}},
\end{equation}
whose lowest eigenvalues are the bound-state energies ($R_{\max}=300$--$400$\,nm, $N=3$--$4\times10^4$).

\paragraph{Optical weight.} A near-field spot carrying orbital angular momentum $m$ couples only to $\ell=m$. The light acts on the bright component of the local exciton, $v_B^2=(1+\cos\theta)/2$ for a fully winding electron and pinned hole, which vanishes in the core. With a Gaussian spot $g_m(R)=(R/s)^{|m|}e^{-R^2/2s^2}$, $s=30$\,nm, and the scattering solution $u_E$ normalized to unit asymptotic amplitude, the optical weight is
\begin{equation}
S_m(E)=\frac{1}{k}\left[\int_0^\infty dR\,\sqrt R\,g_m(R)\,v_B(R)\,u_E(R)\right]^2 .
\label{Seq:optics}
\end{equation}
The factor $1/k$ converts the energy-independent asymptotic normalization into the density of states of the free channel.

\begin{figure}[p]\centering\includegraphics[width=\textwidth]{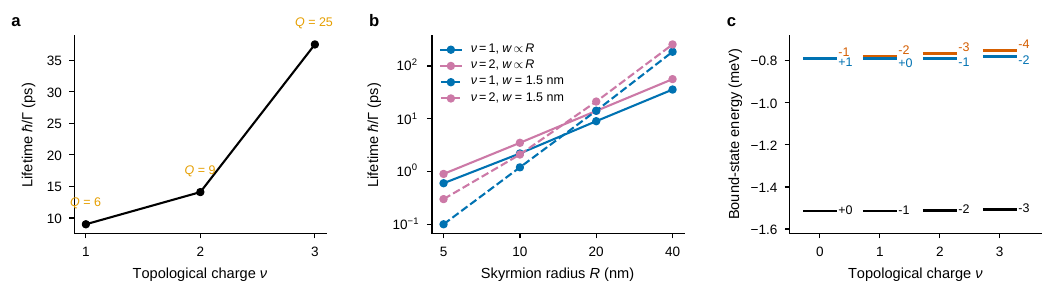}
\caption{\textbf{Trends within the family of 360$^\circ$-wall profiles.} \textbf{a}, Lifetime and quality factor of the $s$-like trapped mode versus topological charge (radius 20\,nm, wall 2\,nm). \textbf{b}, Lifetime versus radius for self-similar skyrmions (wall width $\propto R$) and for a fixed wall width of 1.5\,nm. \textbf{c}, Bound states in an additional exchange well of 2\,meV (radius 20\,nm, wall 1.5\,nm), labelled by $\ell$; the ground state carries $\ell=-\nu$. Energies and lifetimes depend on the wall profile (main text Fig.~\ref{fig:cavity}c).}\label{fig:S7}\end{figure}

Within this family of 360$^\circ$-wall profiles, the lifetime of the $s$-like resonance increases with the topological charge, from 9\,ps for $\nu=1$ to 38\,ps for $\nu=3$ at a radius of 20\,nm, with quality factors between about 6 and 25 (Supplementary Fig.~\ref{fig:S7}a). It also increases strongly with the size of the skyrmion, reaching hundreds of picoseconds for radii of 40\,nm when the wall width is kept fixed (Supplementary Fig.~\ref{fig:S7}b). These trends are useful for design, but we emphasize that they depend on the profile and are not protected by topology. When an exchange well of 2\,meV is added, the resonances become bound states. The ground state then carries $\ell=-\nu$ for all charges considered, and the two excited states with $\ell=-\nu\pm1$, which would be degenerate in an ordinary circular trap, are split by the emergent field (Supplementary Fig.~\ref{fig:S7}c).

\section{Transport and single-exciton occupation}
\label{sn:transport}
\paragraph{Galilean invariance.} For a trap moving at constant velocity, $H=\mathbf P^2/2M+U(\R-\mathbf vt)$, the function
\begin{equation}
\psi(\R,t)=\exp\!\Big[\tfrac{i}{\hbar}\big(M\mathbf v\cdot\R-\tfrac12Mv^2t\big)-\tfrac{i}{\hbar}E_0t\Big]\,\psi_0(\R-\mathbf vt)
\end{equation}
solves the time-dependent Schr\"odinger equation exactly if $\psi_0$ is a bound state of the trap at rest. A trap moving at constant velocity therefore carries its bound state without loss.

\paragraph{Sudden boost.} If the velocity changes instantaneously by $\Delta\mathbf v$, the wavefunction is unchanged in the frame at rest but acquires the phase $e^{iM\Delta\mathbf v\cdot\R/\hbar}$ in the new frame. The probability of remaining bound is $P=|\braket{\psi_0|e^{iq x}|\psi_0}|^2$ with $q=M\Delta v/\hbar$. For an $s$ state, the angular integral gives a Bessel function,
\begin{equation}
P(\Delta v)=\left|\int_0^\infty 2\pi R\,dR\,\rho_0(R)\,J_0(qR)\right|^2,\qquad \rho_0=|\psi_0|^2 .
\label{Seq:boost}
\end{equation}
Any finite acceleration loses less than this bound.

\paragraph{Tunnelling under acceleration.} An acceleration $a$ adds a force $F=Ma$ in the comoving frame. For a triangular barrier, the WKB exponent is $\tfrac43\kappa E_b/F$ with $\kappa=(E_b/t_M)^{1/2}$, so tunnelling becomes appreciable only for $a\gtrsim a^*=\tfrac43\kappa E_b/M$, of order $10^{17}$\,m\,s$^{-2}$ for the parameters of Fig.~\ref{fig:carrier}.

\paragraph{Thermal escape and range.} The escape time is estimated from an Arrhenius law with an attempt frequency given by the level spacing of the trap,
\begin{equation}
\tau_{\rm esc}=\nu_0^{-1}e^{E_b/k_BT},\qquad \nu_0=\frac{E_1-E_0}{h},\qquad E_b=-E_0,
\label{Seq:esc}
\end{equation}
where $E_0$ and $E_1$ are the two lowest $\ell=0$ levels. The transport distance is $L=v\,\min(\tau_{\rm esc},\tau_X)$, with $\tau_X$ the exciton lifetime.

\paragraph{Two excitons in one trap.} Interlayer excitons are dipoles of length $d$ perpendicular to the plane. In the Hartree approximation, two excitons in the ground state of the trap interact with energy
\begin{equation}
U_{\rm dip}=\Big\langle\frac{e^2}{4\pi\varepsilon_0\varepsilon}\Big(\frac{2}{s}-\frac{2}{\sqrt{s^2+d^2}}\Big)\Big\rangle,\qquad s=|\R_1-\R_2|,
\label{Seq:dip}
\end{equation}
averaged over $|\psi_0(\R_1)|^2|\psi_0(\R_2)|^2$ by Monte Carlo sampling, with $d=0.65$\,nm and $\varepsilon=4.5$. The trap accepts a single exciton when $U_{\rm dip}>E_b$.

\paragraph{Results for the parameters of main text Fig.~\ref{fig:carrier}.} At 100\,m/s the loss bound of Eq.~\eqref{Seq:boost} is at most about $10^{-4}$, reached for a skyrmion radius of 40\,nm, and it becomes of order unity only for velocity changes of about 10\,km/s. 

For a binding energy of 1.6\,meV ($V_0=2$\,meV, radius 20\,nm), the escape-limited distance at 100\,m/s is about 50\,$\mu$m at 1.5\,K, 2.4\,$\mu$m at 2\,K and only 0.02\,$\mu$m at 4.2\,K. For 4.5\,meV ($V_0=5$\,meV), it is about 47\,$\mu$m at 4.2\,K, 6\,$\mu$m at 5\,K and 1\,$\mu$m at 6\,K. Since interlayer excitons live for 10--100\,ns, corresponding to 1--10\,$\mu$m at this velocity, thermal escape stops being the limiting factor below about 5\,K for the deeper well. These numbers motivate the design rule $E_b\gtrsim10\,k_BT$. For radii below about 25--27\,nm the dipolar repulsion of Eq.~\eqref{Seq:dip} exceeds the binding energy; the crossover radius depends on $d$ and on the dielectric environment. In the full model, the binding energy for $V_0=2$\,meV falls below $k_BT$ at 4.2\,K for radii below about 12\,nm, so that single occupancy is stable between about 12 and 27\,nm.

\paragraph{Reference model and full model.} Figure~\ref{fig:carrier} of the main text compares a reference model, in which only the exchange well $V$ is kept ($A=W=0$, $\ell=0$), with the full model of Eq.~\eqref{Seq:Ul}, which includes the Berry potential and the geometric barrier ($\nu=1$, ground state $\ell=-1$). In the full model, the attempt frequency is taken from the two lowest levels of the $\ell=-1$ channel. Supplementary Table~\ref{tab:S3} summarizes the binding energies. For radii of 20\,nm and above, the two models differ by at most 10\%, and the escape times by less than a factor of two. For small skyrmions, the geometric barrier strongly reduces the binding: for $V_0=2$\,meV, the exciton is no longer bound at a radius of 8\,nm, and only by 0.06\,meV at 10\,nm. In the full model, the moving texture also produces an emergent electric field. In the frame moving with the skyrmion, it adds the static term $-\hbar\mathbf v\cdot\mathbf a$, of order $\hbar v\nu/R\approx3\,\mu$eV for $v=100$\,m/s and $R=20$\,nm, which is negligible compared with the binding energy.

\begin{table}[h]\centering
\caption{{Binding energy $E_b$ (meV) of an interlayer exciton ($M=0.83\,m_0$, wall 1.5\,nm) in the reference model and in the full model.}}
\begin{tabular}{cccc}\hline
Radius (nm) & $V_0$ (meV) & Reference & Full model\\\hline
10 & 2 & 0.86 & 0.06\\
10 & 5 & 3.39 & 2.81\\
20 & 2 & 1.57 & 1.41\\
20 & 5 & 4.48 & 4.39\\
40 & 2 & 1.87 & 1.84\\
40 & 5 & 4.85 & 4.84\\\hline
\end{tabular}\label{tab:S3}\end{table}

\paragraph{Masses.} For interlayer excitons we use the first-principles band masses of ref.~\cite{kormanyos2015}. Magneto-optical experiments indicate heavier electrons~\cite{goryca2019}, which would increase the exciton mass (Supplementary Table~\ref{tab:S2}). A heavier exciton is bound more strongly in a given trap, so that our estimates of binding and transport are conservative in this respect.

\section{Material parameters}
\label{sn:materials}
Supplementary Table~\ref{tab:S1} collects the parameters of the transition-metal dichalcogenides used in this work, and Supplementary Table~\ref{tab:S2} those of the interlayer excitons used in the transport estimates. Two points deserve attention. First, the threshold for the flux switch of intralayer excitons is set by the measured bright--dark splitting $\Delta_{BD}$, which contains the electron--hole exchange, and not by the bare conduction-band splitting. The two can differ strongly, and even in sign, as the comparison of MoS$_2$ and MoSe$_2$ shows. Second, for interlayer excitons the electron--hole exchange is weak, and the threshold is close to the bare conduction-band splitting of the layer facing the magnet. In the model calculations of the main text we use a representative $\Delta_{BD}=5$\,meV, between the measured values for MoSe$_2$ and MoS$_2$. Since the weight of the dark-exciton ring scales as $(J_c/\Delta_{BD})^2$, it is weaker in MoS$_2$ and stronger in MoSe$_2$ than in these calculations.

\begin{table}[t]\centering
\caption{Spin--orbit splittings and band masses at the $K$ point from first-principles $\mathbf{k}\cdot\mathbf{p}$ fits (HSE/LDA, ref.~\cite{kormanyos2015}), and exciton parameters of hBN-encapsulated monolayers from magneto-optics (refs.~\cite{goryca2019,robert2020}; WSe$_2$ from ref.~27 of ref.~\cite{goryca2019}). $\Delta_{BD}>0$ means the dark exciton lies below the bright one; for MoTe$_2$ the dark exciton was not observed and is presumably above. The conduction-band splitting has opposite sign in W-based compounds, where the lowest-energy exciton is dark. $\Delta$ in the main text is half the splitting.}
\begin{tabular}{lccccccc}\hline
 & $2\Delta_c$ (meV) & $2\Delta_v$ (meV) & $\Delta_{BD}$ (meV), exp. & $m_e$ ($m_0$) & $m_h$ ($m_0$) & $\mu$ ($m_0$), exp. & $r_{1s}$ (nm), exp.\\\hline
MoS$_2$ & 3 & 148 & $+14$ & 0.43--0.46 & 0.54 & 0.275 & 1.2\\
MoSe$_2$ & 22 & 186 & $-1.4$ & 0.49--0.56 & 0.59 & 0.350 & 1.1\\
MoTe$_2$ & 36 & 219 & $<0$ & 0.53--0.62 & 0.66 & 0.36 & 1.3\\
WS$_2$ & $-32$ & 429 & $+40$--$55$ & 0.26--0.35 & 0.35 & 0.175 & 1.8\\
WSe$_2$ & $-37$ & 466 & $+40$ & 0.28--0.39 & 0.36 & 0.20 & 1.7\\\hline
\end{tabular}\label{tab:S1}\end{table}

\begin{table}[t]\centering
\caption{Interlayer excitons used in the transport estimates. Total masses from the band masses of Supplementary Table~\ref{tab:S1} (first-principles) and, in brackets, with the heavier electron masses inferred from magneto-optics~\cite{goryca2019}. Threshold $\Delta_c$ is that of the electron layer facing the magnet.}
\begin{tabular}{lccc}\hline
Heterobilayer & electron / hole layer & $M$ ($m_0$) & $\Delta_c$ (meV)\\\hline
MoS$_2$/WSe$_2$ & MoS$_2$ / WSe$_2$ & 0.83 (1.06) & 1.5\\
MoSe$_2$/WSe$_2$ & MoSe$_2$ / WSe$_2$ & 0.85 (1.2) & 11\\
WS$_2$/WSe$_2$ & WS$_2$ / WSe$_2$ & 0.62 & 16\\\hline
\end{tabular}\label{tab:S2}\end{table}

\section{Equations and parameters behind each figure}
\label{sn:curves}
\noindent For each panel we give the equation that was evaluated and the parameters. Unless stated otherwise: $J_c=1$\,meV, $J_v=0.5$\,meV, $\Delta_\mathrm{BD}=5$\,meV, $\nu=1$, $\varepsilon=0.36$\,meV and the profile of Eq.~\eqref{Seq:profile}.

\paragraph{Figure~\ref{fig:map}.}
\begin{itemize}
%\item \textbf{a}: illustration (Mitsuba 3 rendering); no data.
\item \textbf{b}, shaded regions: binding energy $E_b(J,R)$ of the lowest $\ell=0$ state of $U_0$ in Eq.~\eqref{Seq:Ul} with $A=W=0$, $V_0=2J$, $c=R$, $w=R/5$, $M=m_0$, computed with the finite-volume method of Supplementary Note~\ref{sn:trap} on a $34\times34$ grid in $J\in[0.05,100]$\,meV and $R\in[2,200]$\,nm. Light: $E_b>k_BT$; dark: $E_b>10\,k_BT$ (4.2\,K). The solid lines are the two contours.
\item \textbf{b}, vertical lines: adiabatic onset $J=\hbar^2/(2m_0w^2)$ for $w=1.5$ and 5\,nm (Supplementary Note~\ref{sn:switch}).
\item \textbf{b}, dash-dotted lines: radii at which the lifetime $\hbar/\Gamma$ of the $\ell=-1$ resonance reaches 10 and 100\,ps for a fixed wall $w=1.5$\,nm, by log-interpolation of the values of Supplementary Fig.~\ref{fig:S7}b ($R=18$ and 33\,nm).
\item \textbf{b}, bars: indicative exchange ranges (proximity in direct contact 1--5\,meV; magnetic semiconductors 50--100\,meV).
\item \textbf{c}: $J^*=|\Delta_\mathrm{BD}|/2$ for intralayer excitons (Eq.~\eqref{Seq:Jstar}) with the measured $\Delta_\mathrm{BD}$ of Supplementary Table~\ref{tab:S1}; $J^*=\Delta_c$ of the electron layer for interlayer excitons (Supplementary Table~\ref{tab:S2}).
\end{itemize}

\paragraph{Figure~\ref{fig:bands}.}
\begin{itemize}
%\item \textbf{a}: schematic band edges; no data.
\item \textbf{b}: schematic of Eq.~\eqref{Seq:switch}: path of $\hat{\mathbf b}_i$ for $J<\Delta$ and $J>\Delta$.
\item \textbf{c}: eigenvalues of Eq.~\eqref{Seq:Htau2} versus $R$ for $\tau=\pm1$; $c=5$\,nm, $w=1$\,nm. ``Bright'' and ``dark'' label the branches by their character outside the skyrmion.
\end{itemize}

\paragraph{Figure~\ref{fig:switch}.}
Parameters: $c=10$\,nm, $w=4$\,nm, $J_v=0.5$\,meV, $\Delta_\mathrm{BD}=5$\,meV, $M=1.01\,m_0$, $\varepsilon=0.36$\,meV, partial waves $|\ell|\le8$.
\begin{itemize}
\item \textbf{a}: net flux $\Phi/\Phi_0=A_{\rm reg}(\infty)$ of the bright-connected branch (Supplementary Note~\ref{sn:weak}) for $\tau=\pm1$, at 400 values of $J_c/\Delta_\mathrm{BD}\in[0.04,12]$. Shaded: dark core, $|v_D(0)|^2=1$.
\item \textbf{b}: skew $\langle\sin\theta_s\rangle$ of $|f_{BB}|^2$ (Supplementary Note~\ref{sn:weak}), $\tau=\pm1$, from the exact two-channel solution (circles; closed-channel matching for $K'$ above $J_c/\Delta_\mathrm{BD}=0.54$) and from the single-channel adiabatic problem, Eq.~\eqref{Seq:adradial} (dashed), at 64 ($K$) and 161 ($K'$) values of $J_c/\Delta_\mathrm{BD}\in[0.05,10]$. Increasing the number of partial waves to $|\ell|\le12$ or the matching radius from $c+22w$ to $c+30w$ leaves all curves unchanged (Source Data).
\item \textbf{c}: $\sigma_{\rm el}$ from both solutions.
\item \textbf{d}: $\sigma_{B\to D}/\sigma_{\rm el}$ of the exact solution.
\end{itemize}

%\paragraph{Figure~\ref{fig:switch}.}
%Parameters: $c=10$\,nm, $w=4$\,nm, $J_v=0.5$\,meV, $\Delta_\mathrm{BD}=5$\,meV, $M=1.01\,m_0$, $\varepsilon=0.36$\,meV, partial waves $|\ell|\le8$.
%\begin{itemize}
%\item \textbf{a}: net flux $\Phi/\Phi_0=A_{\rm reg}(\infty)$ of the bright-connected branch (Supplementary Note~\ref{sn:weak}) for $\tau=\pm1$, at 40 values of $J_c/\Delta_\mathrm{BD}\in[0.05,10]$. Shaded: dark core, $|v_D(0)|^2=1$.
%\item \textbf{b}: skew $\langle\sin\theta_s\rangle$ of $|f_{BB}|^2$ (Supplementary Note~\ref{sn:weak}), $\tau=+1$, from the exact two-channel solution (circles) and from the single-channel adiabatic problem, Eq.~\eqref{Seq:adradial} (squares).
%\item \textbf{c}: $\sigma_{\rm el}$ from both solutions.
%\item \textbf{d}: $\sigma_{B\to D}/\sigma_{\rm el}$ of the exact solution.
%\end{itemize}

\paragraph{Figure~\ref{fig:cavity}.}
Parameters: $\nu=1$, $c=20$\,nm, $w=2$\,nm, $M=1.01\,m_0$, $V_0=0$, fully winding exciton, Eq.~\eqref{Seq:AWfull}.
\begin{itemize}
\item \textbf{a}: $W(R)$ (grey) and $t_M(\ell-A)^2/R^2+W$ for $\ell=-1,0,+1$ (the two-dimensional form of Eq.~\eqref{Seq:Ul}, without the $-\tfrac14$ of the radial equation). Dotted line: resonance energy 0.43\,meV.
\item \textbf{b}: $\tau_W(E)$, Eq.~\eqref{Seq:tauW}, for 1571 energies in $[0.03,1.6]$\,meV, from the radial equation integrated to $R_{\max}=120$\,nm: skyrmion, $\ell=-1$; no-flux control ($A=0$), $\ell=0$; skyrmionium with $c_1=10$, $c_2=20$\,nm, $w=2$\,nm, $\ell=0$ and $\ell=-1$.
\item \textbf{c}: $\tau_W(E)$ for $\ell=-1$ with $w=1$, 2 and 4\,nm, and for the linear wall with $w=2$\,nm. Resonances: 0.48, 0.43 and 0.34\,meV with lifetimes $\tau_W^{\max}/4=28$, 9.0 and 3.4\,ps; linear wall 0.51\,meV, 49\,ps.
\item \textbf{d}: $S_m(E)/S_m(E_{\min})$ from Eq.~\eqref{Seq:optics} for $m=-1,0,+1$, $s=30$\,nm, $E_{\min}=0.03$\,meV, 600 energies, with $u_E$ normalized to unit asymptotic amplitude of $u$. An independent calculation from the eigenstates of a large box ($R_{\max}=700$\,nm), $S_m(E_n)\propto|\langle g_mv_B|n\rangle|^2/\Delta E_n$, reproduces the curves to within 10\% and the positions of all zeros.
\end{itemize}

%\paragraph{Figure~\ref{fig:cavity}.}
%Parameters: $\nu=1$, $c=20$\,nm, $w=2$\,nm, $M=1.01\,m_0$, $V_0=0$, fully winding exciton, Eq.~\eqref{Seq:AWfull}.
%\begin{itemize}
%\item \textbf{a}: $W(R)$ (grey) and $U_\ell(R)$ of Eq.~\eqref{Seq:Ul} for $\ell=-1,0,+1$. Dotted line: resonance energy 0.42\,meV.
%\item \textbf{b}: $\tau_W(E)$, Eq.~\eqref{Seq:tauW}, for 110 energies in $[0.03,1.6]$\,meV: skyrmion, $\ell=-1$; no-flux control ($A=0$), $\ell=0$; skyrmionium with $c_1=10$, $c_2=20$\,nm, $w=2$\,nm, $\ell=0$ and $\ell=-1$.
%\item \textbf{c}: $\tau_W(E)$ for $\ell=-1$ with $w=1$, 2 and 4\,nm, and for the linear wall with $w=2$\,nm.
%\item \textbf{d}: $S_m(E)/S_m(E_{\min})$ from Eq.~\eqref{Seq:optics} for $m=-1,0,+1$, $s=30$\,nm, $E_{\min}=0.03$\,meV.
%\end{itemize}

\paragraph{Figure~\ref{fig:carrier}.}
Parameters: interlayer exciton, $M=0.83\,m_0$, exchange well $V_0=2$\,meV, wall $w=1.5$\,nm, $c=R$. Solid lines: reference model, $A=W=0$, $\ell=0$. Dashed lines: full model, Eq.~\eqref{Seq:Ul} with $A$ and $W$ from Eq.~\eqref{Seq:AWfull}, $\nu=1$, ground state $\ell=-1$, attempt frequency from the two lowest $\ell=-1$ levels.
\begin{itemize}
\item \textbf{a}: $1-P(\Delta v)$ from Eq.~\eqref{Seq:boost} for $R=10$, 20 and 40\,nm, with the ground state from the finite-volume solver ($R_{\max}=300$\,nm, $N=3\times10^4$).
\item \textbf{b}: $L=v\,\tau_{\rm esc}(T)$ with $v=100$\,m/s and Eq.~\eqref{Seq:esc}, for $R=20$\,nm and $V_0=2$ and 5\,meV, which give $E_b=1.6$ and 4.5\,meV. Bands: $v\tau_X$ for dark intralayer excitons ($\tau_X=0.1$--1\,ns) and interlayer excitons ($\tau_X=10$--100\,ns).
\item \textbf{c}: $U_{\rm dip}$ from Eq.~\eqref{Seq:dip} ($1.5\times10^5$ samples) and $E_b$, for $R=8$, 10, 14, 20, 28, 40 and 55\,nm. Dotted line: $k_BT$ at 4.2\,K. Shaded: full-model radii with $E_b>k_BT$ and $U_{\rm dip}>E_b$ (single occupancy).
\end{itemize}

\paragraph{Supplementary Figures.}
\begin{itemize}
\item \textbf{S1}: $GW$--BSE data plotted directly. (a) $\mathrm{Im}\,\alpha(\omega)$ with (BSE) and without (independent quasiparticles) electron--hole interaction, normalized to the maximum in the plotted range; quasiparticle gap $=\min_k[E^{\rm QP}_{27}(k)-E^{\rm QP}_{26}(k)]$. (b) Oscillator strengths of the lowest states, energies relative to the bright A exciton. (c) Exciton density; no model.
\item \textbf{S2}: structureless spinor model, $H=(\mathbf P-\hbar\mathbf a)^2/2\mu_X+W$, with $\mu_X=5.6\,m_0$, $\Theta(r)=2\arctan[\cosh(c/(w/2))/\sinh(r/(w/2))]$, $c=4.9$\,nm, $w=1.0$\,nm, $A=-(1-\cos\Theta)/2$, $W=\tfrac{t_X}{4}[\Theta'^2+\sin^2\Theta/r^2]$, $k=3.3$, 9.9 and 16.5\,$\mu$m$^{-1}$, $|\ell|\le6$. (a) $|f(\theta_s)|^2=|\sum_\ell(S_\ell-1)e^{i\ell\theta_s}|^2/(2\pi k)$; dotted: $4/(k\theta_s^2)$. (b) $\gamma(\theta_s)$.
\item \textbf{S3}: exact two-channel solution of Supplementary Note~\ref{sn:weak} with $w=1$\,nm. (a) $\gamma(\theta_s)$ for $c=10$\,nm, $\tau=\pm1$. (b) Skew versus $c=1,\dots,14$\,nm. (c) Skew versus $\varepsilon=0.1$--3\,meV for $c=3$ and 10\,nm. (d) $\sigma_{B\to D}/\sigma_{\rm el}$ versus $c$.
\item \textbf{S4}: (a) $E_K-E_{K'}$ of the bright branch and (b) bright weight of the dark branch in $K$, from Eq.~\eqref{Seq:Htau} with the averaged fields of Supplementary Note~\ref{sn:weak} ($c=5$\,nm, $w=1$\,nm, $a_B=1.5$\,nm, $m_e=0.47$, $m_h=0.54$). (c) $E_b$ of the lowest $\ell=0$ state in $V=-2|J_c-J_v|(1-\cos\theta)/2$, $w=1$\,nm, $M=1.01\,m_0$. (d) Eq.~\eqref{Seq:ring}.
\item \textbf{S5}: first-principles $J_{ij}$ and $|\mathbf D_{ij}|$ of the Heisenberg model $H=-\sum_{ij}J_{ij}\mathbf S_i\cdot\mathbf S_j-\sum_{ij}\mathbf D_{ij}\cdot(\mathbf S_i\times\mathbf S_j)$ from KKR; no model equation.
\item \textbf{S6}: (a) Supplementary Note~\ref{sn:pdfe}. (b) Core valley reversal $-[E_K-E_{K'}](0)/[2(J_c-J_v)]$ and maximal ring weight, both relative to a large skyrmion ($c=20$\,nm, $w=0.75$\,nm), for $a_B=1.0$ and 1.5\,nm, using the profiles of (a) with $w=w_R/2$. (c) Map at 1.15\,T, $a_B=1.5$\,nm, with contours of the bright weight of the dark branch. (d) $\sigma_{B\to D}/\sigma_{\rm el}$ from Supplementary Note~\ref{sn:weak} at each field.
\item \textbf{S7}: (a) $\hbar/\Gamma$ and $Q$ of the $\ell=-\nu$ resonance for $\nu=1,2,3$, $c=20$\,nm, $w=2$\,nm. (b) $\hbar/\Gamma$ versus $c$ for self-similar skyrmions ($w=c/10$) and for $w=1.5$\,nm. (c) Bound states of Eq.~\eqref{Seq:Ul} with $V_0=2$\,meV, $c=20$\,nm, $w=1.5$\,nm, labelled by $\ell$.
\end{itemize}

\end{document}